\documentclass[12pt,subeqn,a4paper]{article}

\usepackage{amssymb,amsmath,amsfonts,amsthm, amscd, mathrsfs,helvet}
\usepackage{bm}
\usepackage{graphicx,verbatim}
\usepackage{psfrag}
\usepackage[all]{xy}
\usepackage{color}
\usepackage{epsfig}
\input epsf

\def\bes{\cite{bes}}
\def\bhl{\cite{bhl}}
\def\hm{\cite{hm}}
\def\be{\begin{equation} }
\def\ee{\end{equation} }
\def\la#1{\label{#1}}
\def\bea{\begin{eqnarray} }
\def\eea{\end{eqnarray}}
\def\nref#1{(\ref{#1})}

\begin{document}

\begin{titlepage}
\begin{flushright}
{\bf March 2007} \\
%DAMTP-\\
%SWAT-\\
hep-th/0703104 \\
\end{flushright}
\begin{centering}
\vspace{.2in}
 {\large {\bf On the Singularities of the Magnon S-matrix  }}

\vspace{.3in}

Nick Dorey$^{1}$, Diego M. Hofman$^{2}$ and Juan Maldacena$^{3}$ \\
\vspace{.1 in}
$^{1}$DAMTP, University of Cambridge, Cambridge CB3 0WA, UK. \\
$^{2}$ Joseph Henry Laboratories, Princeton
University, Princeton, NJ 08544, USA. \\
$^{3}$  Institute for Advanced Study, Princeton NJ 08540, USA.
\vspace{.2in} \\
%and \\
%\vspace{.2in}
%
%
\vspace{.4in}
{\bf Abstract} \\

\end{centering}
We investigate the analytic structure of the magnon S-matrix in the
spin-chain description of planar ${\cal N}=4$ SUSY
Yang-Mills/$AdS_{5}\times S^{5}$ strings. Semiclassical analysis
suggests that the exact S-matrix must have a large family of poles
near the real axis in momentum space.
In this article we show that these are double poles corresponding
to the exchange of pairs of BPS magnons.  Their
locations in the complex plane are uniquely fixed by the known
dispersion relation for the BPS particles. The locations precisely
agree with the recent conjecture for the $S$ matrix by
Beisert, Hernandez, Lopez, Eden and Staudacher
(hep-th/0609044 and hep-th/0610251).
These poles do not signal the presence of new bound states.
In fact, a certain non-BPS localized classical solution, which was
thought to give rise to new bound states,
can actually decay into a pair of BPS magnons.

%\vspace{.05in}
%\baselineskip=.3in
\end{titlepage}

\section{Introduction}
\paragraph{}
In 't Hooft's large $N$ limit, a gauge theory is reduced to the
sum of planar diagrams. These diagrams give rise to a two
dimensional effective theory which is supposed to be the
worldsheet of a string. A great deal of effort has been devoted
recently to studying
 planar   ${\cal N}=4$ super
Yang Mills, culminating in a conjecture  for the exact $\lambda =
g^2 N$ dependence of some quantities \cite{bhl,bes} (see also
\cite{es}). This has been done using the assumption of exact
integrability plus other reasonable assumptions. The quantity that
has been conjectured is
the exact  $S$ matrix describing the scattering of worldsheet
excitations. Let us briefly describe how this quantity is defined
in the gauge theory \cite{ssmatrix}. In ${\cal N}=4$ super Yang
mills we have an $SO(6)$ R-symmetry. We pick an $SO(2)$ subgroup
generated by $J = J_{56}$. We denote by $Z$ the scalar field that
carries charge one under $J$. We then consider singe trace local
operators with very large charge, $J\to \infty$, and conformal
dimension, $\Delta$,
 close to $J$, so that $\Delta - J$
is finite in the limit. In this limit the local operator contains a
large number of fields $Z$ and a finite number of other fundamental
fields. These $Z$ fields form a sort of one dimensional lattice on
which the other fields propagate. It turns out that we can describe
all the other fields in terms of a set of 8 bosonic and 8 fermionic
fundamental fields \cite{bsbethe} . The 8 bosonic ones are four of
the scalar fields that are not charged under $J$ and the four
derivatives of $Z$, $\partial_\mu Z$. We can view the different
fields at each site as a generalized spin. Therefore, we are dealing
with a generalized spin chain. For this reason the fundamental
excitations are often called ``magnons''. The symmetry algebra
acting on an infinite chain is enhanced in such a way that the
fundamental excitations (or magnons) are BPS for all values of their
momenta \cite{beiserts}.
\paragraph{}
 This symmetry also completely constrains the matrix structure of the
 $ 2 \to 2$ scattering amplitude \cite{beiserts}. Only the overall
 phase is undetermined.
  The phase is constrained by a
crossing symmetry equation \cite{janik,AF}. Recently, an expression for this
phase, sometimes called ``dressing factor'' \cite{AFS},  was proposed in
\cite{bhl,bes}. The proposed phase is a non-trivial function of
 the 't Hooft coupling $\lambda$ and is supposed to be valid for all
 values of this parameter.
The function also depends on the two momenta of the particles
$p_1,p_2$ and can be analytically continued to complex values of
these variables. As we do so, we encounter poles and branch points.
In this article we explore the physical meaning of some poles that
appear when we perform this analytic continuation. For $S$ matrices
in any dimension, simple poles are typically associated
 to on shell intermediate states. In fact, such simple
poles appear in some of the matrix elements of the full matrix $S$
and are independent of the dressing factor. They were interpreted
as BPS bound states for single magnon excitations in
\cite{doreymagone}.
 It turns out that one
can have BPS bound states of $n$ fundamental magnons for any
positive integer $n$.
 Further poles were necessary in order to account for
branch cuts in the scattering of magnons in the classical limit
considered in \cite{hm}. In the proposal of \cite{bhl,bes}
these turn out to be double poles. Such double poles do not arise
from bound states and look a bit puzzling at first sight.  However,
similar double poles appear in the $S$ matrix for sine Gordon theory
and their physical origin was elucidated by Coleman and Thun
\cite{ct} (see \cite{pdorey} for  a recent review on two dimensional
S-matrices).
 In short, they arise from physical processes where the elementary
particles exchange pairs of particles rather than a single particle.
In higher dimensions such processes give rise to branch cuts, but
they give double poles in two dimensions. The precise position of
these double poles depends on the dispersion relation for the
exchanged particles. If we assume that the exchanged particles are
the BPS magnon bound states discussed above we find that the double
poles appear precisely where the conjecture of \cite{bhl,bes} predicts them.
Thus, this computation can be viewed a a check of their proposal, since
it has singularities where (with hindsight) one would have expected
them.
\paragraph{}
More generally, the existence
 of the singularities dictated by the spectrum (and no others in the
 physical region) provides
extra constraints on the S-matrix beyond those of unitarity, crossing
 and factorizability. As in relativistic models \cite{Zam}, these
constraints help to remove ambiguities associated with the
homogeneous solutions of the crossing equations. Optimistically, one
can hope that this type of reasoning might provide a physical basis
for selecting the conjectured S-matrix of \cite{bes} from the many
possible solutions to the crossing equation.
\paragraph{}
In \cite{hm} some non-BPS localized classical solutions were found
and it was suggested that they could correspond to non-BPS bound
states. If true this would mean that the S-matrix should have
single poles rather than double poles. These solutions move in an
$R \times S^2$ subspace of the full $AdS_5 \times S^5$ theory and
within this subspace they are stable.
 Here we show that these
solutions can decay once they are embedded in $AdS_5 \times S^5$.
In fact these solutions can be viewed as a non-BPS superposition
of two coincident BPS magnons with large $n$ and the same
momentum. The decay of the solutions corresponds to these two
BPS magnons moving away from each other. Our results suggest
that the only single-particle asymptotic states of the theory are
the tower of BPS boundstates described in
\cite{doreymagone,doreymagtwo,doreymagthree,doreymagfour}.
\paragraph{}
This article is organized as follows. In Section two we review
Coleman and Thun's explanation for the double poles. In Section
three we use the magnon dispersion relation to give the location
of the double poles. In Section four we discuss the localized
classical solutions found in \cite{hm,sv} and their
relation to the double poles. We also discuss a slow relative
velocity limit, or non-relativistic limit, where the system seems
to simplify. In particular we present a toy model, in this limit,
which reproduces some of the features of the full theory. In
Section five we compute the location of the double poles in the
recent conjecture of \cite{bhl,bes}. The derivation of a useful
integral representation of the conjectured phase is given in an
Appendix. Finally, we include two other Appendices where we
present some diagrams related to the double poles and where we
include explicitly the classical solutions of \cite{sv}.

\section{S-matrix Singularities}
\begin{figure}
 \centering
 \includegraphics[width=100mm] {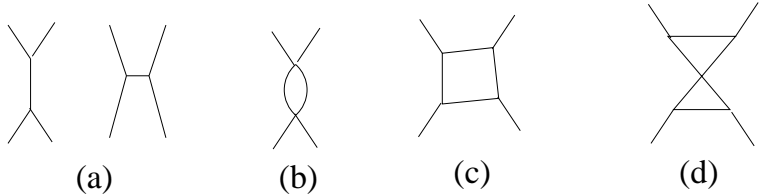}
\caption{Diagrams associated to singularities of the S-matrix. }
\label{diagrams}
\end{figure}

\paragraph{}
Since the S-matrix is a physical observable, its singularities
should have a physical explanation.
 In fact, S-matrix
singularities arise  when we produce on shell particles that
propagate over long distances or long times. Thus the singularities
are interpreted as an IR phenomenon associated to the propagation of
particles.
\paragraph{}
Let us consider a $2 \to 2$ scattering process.  The simplest and
most familiar  example is a single pole. These poles arise when an
intermediate on shell particle is produced in the collision.
See Figure \ref{diagrams}(a).
\paragraph{}
If we have more than one particle becoming on shell we can have
other types of singularities. The simplest example is a two particle
threshold, where we can start producing a pair of intermediate
on-shell particles, see Figure \ref{diagrams}(b). For our purposes
we will need to consider  more complicated cases. This problem was
studied by Landau \cite{landau} who found the general rules for
locating the singularities. The singularities   correspond to
spacetime graphs where the vertices are points in spacetime,
representing local interaction regions. Lines are on shell particles
with particular momenta $p^\mu$ and the spacetime momenta are
conserved at the vertices. In addition, the four momentum of the
line connecting two vertices obeys $x_2^\mu - x_1^\mu = \alpha p^\mu
$, with $\alpha > 0$ (there is one $\alpha$ for each internal line).
Thus, the singularity is associated to the physical propagation of
these on shell particles. Notice that this condition implies that if
the energy $p^0$ is positive, then $x_2^0> x_1^0$, so that the
particle is going forwards in time. More details can be found in
\cite{cutkosky,smatrixbooks}.
 The type of singularity that we obtain
depends on the dimension, since we generally have integrals over
loop momenta. We typically encounter branch points\footnote{The
discontinuity across the branch cut emanating from it can be
evaluated using the Cutkosky rules\cite{cutkosky}.}.  For the box
diagram in Figure \ref{diagrams}(c) we  have a $D$ dimensional
loop integral and one mass shell delta function for each line. We
expect a divergence in $D < 4$ dimensions and branch points for
$D\geq 4$. In two dimensions, $D=2$,  we naively get the square of
a delta function at zero. A more careful analysis reveals that we
get a double pole \cite{ct,smatrixbooks}. Other graphs, such as
the two loop diagram in Figure \nref{diagrams}(d),
 also give rise to double poles. In two dimensions we get double poles
 in diagrams with
 $L$ loops and $N$ internal lines when $ N - 2 L =2 $.
\paragraph{}
So far, we were assuming that the singularities
arise for real values of the external momenta.   On the other hand,
it is often the case that the singularities only arise when we
perform an analytic continuation in the external momenta. The
spacetime points discussed above now live in a complexified
spacetime. If we are interested in the ``first'' singularity that
appears as we move away from the physical sheet,  it is sufficient
to consider momenta which are in ``anti-Euclidean'' space where
$p^0$ is real and $p^i$ are purely imaginary \cite{smatrixbooks}. As
we go through branch cuts we can encounter additional singularities
which relax some of the rules discussed above.
In particular, we can relax the condition that the $ \alpha$'s for
each line are positive. Here we will be interested in the first
singularity that arises and not on the ones that are on other sheets.
In two dimensions, the energy and momentum conservation conditions and
the on shell conditions for all internal lines determine the energies
and momenta up to some discrete options. At any of these values we
will have a singularity in some sheet.
If we are interested in the ``first'' or physical  sheet singularities
then we need to impose a the further condition that the $\alpha_i$
discussed a above are positive, or equivalently, that
the graph in ``anti-Euclidean'' space closes. This selects a subset of all the
discrete solutions to the energy and momentum conservation conditions.
\paragraph{}
In our case, we do not have a relativistic theory, so we do not
know a priory how much we can analytically continue the amplitudes
and expect to find a physical explanation for all singularities.
However, since the origin of the singularities is an IR
phenomenon we expect
 that the discussion should also hold for spin chains.
Understanding precisely the whole physical region is beyond the
scope of this paper. Here we will perform the analytic continuation
only within a very small neighborhood of the real values   looking
for the ``first'' singularities. As we explained above, after we
impose the momentum and
energy conservation conditions we will be left with discrete
possible solutions. We should
select among them to find the singularities that are on the physical sheet.
We are not going to give a general rule, as was given for relativistic
theories. Instead we are going to analyze mainly two cases in this
paper. In the first case we will be
scattering particles in the near plane wave region where particles are
approximately relativistic and we can apply the relativistic rules in
order to check whether the singularity is on
the first sheet or not. The second case is when the particles we
scatter are in the giant magnon region. In this case we can use a slow
motion, non relativistic approximation plus some physical arguments
regarding the interpretation of the solutions in order to select among
different solutions.

\section{Poles from physical processes}
\paragraph{}
As discussed above, the singularities of the S-matrix correspond
to different on-shell intermediate states. In particular, the
location and nature of the singularities are essentially
determined once the spectrum of the theory is known. To understand
the poles of the magnon S-matrix we therefore begin by reviewing
the spectrum of the ${\cal N}=4$ spin chain.
\paragraph{}
The fundamental excitations of the spin chain are the magnons
themselves, which lie in a sixteen-dimensional BPS representation
of the unbroken $SU(2|2)\times SU(2|2)$ supersymmetry. The closure
of the SUSY algebra on this multiplet uniquely determines
\cite{beiserts} the magnon dispersion relation to be \cite{bds} (see also
\cite{Dispersion}),
\begin{equation}
E=\Delta-J=\sqrt{1+16g^{2}\sin^{2}\left(\frac{p}{2}\right)}
\end{equation}
where $g=\sqrt{g^{2}_{YM}N}/4\pi$. In addition, any number of
elementary magnons can form a stable boundstate. The $n$-magnon
boundstate also lies in a BPS representation of supersymmetry (of
dimension $16n^{2}$). The theory therefore contains an infinite
tower of BPS states labelled by a positive integer $n$. The exact
dispersion relation for these states is again fixed by
supersymmetry to be \cite{doreymagone,doreymagfour},
\begin{equation}
E=\Delta-J=\sqrt{n^{2}+16g^{2}\sin^{2}\left(\frac{p}{2}\right)}
\label{dsp1}
\end{equation}
The existence of these states can be confirmed both in the gauge
theory spin-chain and in the string world-sheet theory for
appropriate values of the coupling \cite{doreymagone}. It
remains possible that there are additional boundstates in the
theory which are not BPS. However we will see that we have no
reason to introduce them, at least if one believes in the recently
conjectured $S$ matrix in \cite{bhl,bes}. In fact, the poles in this
proposed result  can be   accounted for by thinking about physical
processes involving  the set of BPS states described above.

\subsubsection*{Kinematics}
\paragraph{}
The dispersion relation (\ref{dsp1}) for an arbitrary BPS state is
conveniently written in terms of complex spectral parameters,
\begin{eqnarray}
X^{\pm} &= &e^{\pm i p/2} \,
\frac{\left(n+\sqrt{n^{2}+16g^{2}\sin^{2}\left(\frac{p}{2}\right)}\right)}
{4g\sin \left(\frac{p}{2}\right)}
\end{eqnarray}
which obey the constraint,
\begin{eqnarray}
\left( X^{+}+\frac{1}{X^{+}}\right)-\left(
X^{-}+\frac{1}{X^{-}}\right) & = & \frac{in}{g} \label{const}
\end{eqnarray}
In terms of these parameters, the particle energy and momentum are
given by,
\begin{eqnarray}
E(X^{\pm})=\frac{g}{i}\left[ \left(
X^{+}-\frac{1}{X^{+}}\right)-\left(
X^{-}-\frac{1}{X^{-}}\right)\right] , & \qquad{} & p(X^{\pm})=
\frac{1}{i}\log\left(\frac{X^{+}}{X^{-}}\right) \label{ep}
\end{eqnarray}
These quantities are real provided $X^{+}=(X^{-})^{*}$. More
generally we will consider an analytic continuation of the
kinematic variables where the reality condition is relaxed (but
the constraint (\ref{const}) is maintained).
\paragraph{}
For magnons belonging to an $SU(2)$ subsector, the integer $n$
appearing in the constraint (\ref{const}) corresponds to a
conserved $U(1)$ R-charge. It is convenient to view  (\ref{const})
as giving the charge as a function of the spectral parameters
$n(X^\pm)$.
%\begin{equation}
%n(X^{\pm})=\frac{g}{i}\left[
%\left( X^{+}+\frac{1}{X^{+}}\right)-\left(
%X^{-}+\frac{1}{X^{-}}\right)\right]
%\end{equation}
The velocity of the particle in
appropriately-normalised\footnote{The
  normalisation is chosen so that the velocity of light is unity
in the plane wave limit discussed below. With this normalization
$p$ generates translations in $x/(2g)$. } worldsheet coordinates
  $(x,t)$ is given as,
\begin{equation}
v=\frac{dx}{dt}=\frac{1}{2g}\frac{dE}{dp}=\frac{X^{+}+X^{-}}{1+X^{+}X^{-}}=
\frac{2g\sin(p)}{\sqrt{n^{2}+16g^{2}\sin^{2}\left(\frac{p}{2}\right)}}
\label{vel}
\end{equation}
and we also define the rapidity parameter,
\begin{eqnarray}
u(X^{\pm}) & = & \frac{1}{2}\left[
\left(X^{+}+\frac{1}{X^{+}}\right)+
\left(X^{-}+\frac{1}{X^{-}}\right)\right] \nonumber \\
 & =& \frac{1}{2g}\cot\left(\frac{p}{2}\right)
\sqrt{n^{2}+16g^{2}\sin^{2}\left(\frac{p}{2}\right)}
\end{eqnarray}
\paragraph{}
In the strong coupling limit $g\gg 1$ there are two distinct
kinematic regimes which will be of interest. The first is the
Giant Magnon regime where the conserved momentum $p$ is held fixed
as $g\rightarrow \infty$.  We then have
\begin{equation} \la{gg1}
E\simeq 4g  \sin\left(\frac{p}{2}\right)  ~,~~~~~~X^+ \sim { 1
\over X^-} \sim e^{ i p/2} ~,~~~~u \sim 2 v \sim 2 \cos {p
\over 2}
\end{equation}
These excitations correspond to classical solutions of the
worldsheet theory.
\paragraph{}
The second regime of interest is that of the plane wave limit
where we take $p\rightarrow 0$ as $g\rightarrow \infty$ with
$k=2gp$ held fixed. In this case we recover the familiar
relativistic relations
\begin{equation}
E \simeq \sqrt{n^{2}+k^{2}} ~, ~~~~v = { k \over \sqrt{n^2 + k^2}
} ~,~~~~~~~X^+ \sim X^- = { n + E \over k} \sim { n + \sqrt{ n^2 +
k^2 } \over k } \label{pp1}
\end{equation}
%
%
%
%Correspondingly, the velocity and rapidity become,
%\begin{eqnarray}
%v=\frac{dE}{dk}\simeq \frac{k}{\sqrt{n^{2}+k^{2}}},  & \qquad{} &
%u\simeq \frac{2}{k}\sqrt{n^{2}+k^{2}} ~,~~~~~~X^\pm \sim X^- \sim { n + E \over k} = { n + \sqrt{n^2 + k }
%\over k }
%\label{pp2}
%\end{eqnarray}
%Comparing the above formula we see that the allowed ranges of the
%rapidity $u$ are complimentary in the two regimes. In the Giant
%Magnon regime we always have $|u|\leq 2$ while plane wave
%excitations always have $|u|\geq 2$.
These two regimes\footnote{The two
regimes in question are connected by  a third, studied in
\cite{ms,kloseloop} where $X^\pm \sim 1 $ (or $X^+ \sim X^- \sim
-1$).}, \nref{gg1} and  \nref{pp1},  amount to
two ways of solving the constraint \nref{const} for large $g$, by
setting $X^+ \sim 1/X^{-}$ or $X^+ \sim X^-$.

\subsubsection*{S-matrix definitions and conventions }
\paragraph{}

For simplicity, we will consider scattering of two states in the
$SU(2)$ subsector, since the scattering in the other sectors is
determined once we know the result in the $SU(2)$ subsector
\cite{beiserts}. Of course, we will allow intermediate states to
be completely general. We can define the  $S$ matrix (just a
complex number in this case) by looking at the wavefunction for
two magnons and writing it as
  \be
\la{wfexp} \Psi(x_1,x_2) = e^{ i p_1 x_1 + p_2 x_2} +
S(p_1,p_2) e^{ i p_1 x_2 + i p_2 x_1 } ~,~~~~x_1 \ll x_2 \ee And
$\Psi(x_1,x_2) = \Psi(x_2,x_1)$ for $x_1 \gg x_2$, since we have
two identical bosons. We see that we can view the first term as
the incoming wave and the second as the reflected way if
$v_1>v_2$. We will  be interested in analytically continuing in
the external momenta $p_i$. We write then
\be
\la{momep}
p_{1}=p+iq ~,~~~~~~~~p_2 = p - i q
 \ee
We will see that the part of the wavefunction depending on the relative coordinate
$x = x_1 -x_2$ has the form $\Psi \sim e^{ - q x } + S(1,2) e^{ q x} $.
Thus we see that the first term diverges as $x \to -\infty$ if $q>0$. We set
boundary conditions on the non-normalizable piece of the wavefunction, by saying that the
coefficient of the exponential is one, and we ``measure'' the coefficient of the other which
is the $S$ matrix. We then see that a pole of $S$ (with $q>0$) can correspond to a bound
state.

This $S$ matrix can be written as
\begin{equation}
S\left(X_{1}^{\pm},X_{2}^{\pm}\right)= \sigma^{-2}
\left(X_{1}^{\pm},X_{2}^{\pm}\right)\,
\mathcal{S}^{-1}_{BDS}\left(X_{1}^{\pm},X_{2}^{\pm}\right)
\label{smat}
\end{equation}
where the inverse factors originate from the fact that the conventions
for defining the $S$ matrix in the recent literature are the opposite from ours
\footnote{ Our $\sigma$ is the
same as the $\sigma$ in \cite{bes,bhl} and similarly, our $S_{BDS}$ is
the same S-matrix appearing in \cite{bds}.}.
The second factor in \nref{smat} was originally written in \cite{bds} and is
\begin{equation} \label{bdsfactor}
\mathcal{S}^{-1}_{BDS}\left(X_{1}^{\pm},X_{2}^{\pm}\right)=
\frac{u(X^{\pm}_{1})-u(X^{\pm}_{2})-i/g}
{u(X^{\pm}_{1})-u(X^{\pm}_{2})+i/g}=\frac{(X_{1}^{-}-X_{2}^{+})
(1-1/X_{1}^{-}X_{2}^{+})}{(X_{1}^{+}-X_{2}^{-})
(1-1/X_{1}^{+}X_{2}^{-})}
\end{equation}
The quantity
$\sigma(X_{1}^{\pm},X_{2}^{\pm})$ is known as the dressing factor and
we will discuss the  explicit proposal in \cite{bhl,bes}  in Section 5 below.
For the moment we will simply use the that fact that the proposed
dressing factor has no poles or zeros when $u_1-u_2 = \pm i$ on the
main branch that is closest to physical values.

\subsection{Simple poles }

\paragraph{}
The singularities of the S-matrix
correspond to spacetime diagrams where each particle is on-shell.
In general, we will analytically continue to complex values of the
momenta and energies, and thus each particle corresponds to a
line in complexified Minkowski space. Particle lines can meet at
three-point vertices which conserve charge, energy and momentum.
We will now analyze the allowed vertices.
\begin{figure}
\centering
\psfrag{t}{\footnotesize{$t$}}
\psfrag{a}{\footnotesize{$a)$}}
\psfrag{b}{\footnotesize{$b)$}}
\psfrag{x}{\footnotesize{$X^{\pm}$}}
\psfrag{y}{\footnotesize{$Y^{\pm}$}}
\psfrag{z}{\footnotesize{$Z^{\pm}$}}
\psfrag{v}{\footnotesize{$Y^{\pm}$}}
\psfrag{w}{\footnotesize{$X^{\pm}$}}
\psfrag{u}{\footnotesize{$Z^{\pm}$}}
\includegraphics[width=100mm]{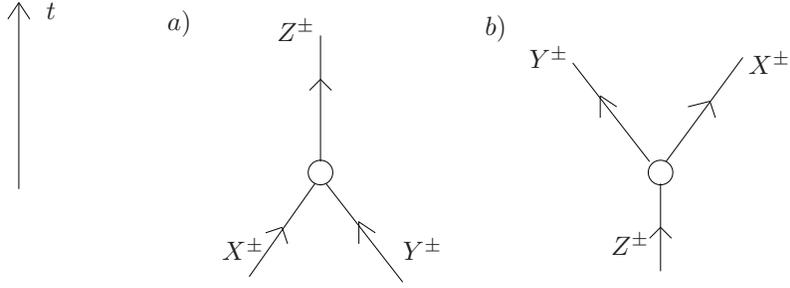}
\caption{Three-point vertices.}
\label{Afig6}
\end{figure}
\paragraph{}
We begin by considering a three point vertex corresponding to
the creation of a
BPS particle with spectral parameters $Z^{\pm}$ from two others with
parameters $X^{\pm}$ and $Y^{\pm}$ as shown in Figure \ref{Afig6}a. If all
particles belong to the same $SU(2)$ subsector, then each carries a
conserved $U(1)_{R}$ charge $Q=n$.
In this case, the
conservation laws for energy, momentum and charge read,
\begin{eqnarray}
E(X^{\pm})+E(Y^{\pm}) & = & E(Z^{\pm}) \nonumber \\
p(X^{\pm})+p(Y^{\pm}) & = & p(Z^{\pm})   ~~~~{\rm mod}(2 \pi) \label{conseqd} \\
n(X^{\pm})+n(Y^{\pm}) & = & n(Z^{\pm}) \nonumber
\end{eqnarray}
The same equations must also hold for the vertex corresponding to the
time-reversed process shown in Figure \ref{Afig6}b. As the dependence of
each conserved quantity on the spectral parameters is of the form
$f(X^{+})-f(X^{-})$ for some function $f$,
it is straightforward to solve the conservation
equations. There are two inequivalent solutions,
\begin{eqnarray}
{\mathbf{\alpha}:} \qquad{} \begin{array}{c} X^{+}=Z^{+} \\
  X^{-}=Y^{+} \\ Y^{-}=Z^{-}
\end{array}  & \qquad{} & {\mathbf{\beta}:}
\qquad{} \begin{array}{c} X^{+}=Y^{-} \\
  X^{-}=Z^{-} \\ Y^{+}=Z^{+}
\end{array}
\label{twochoices}
\end{eqnarray}
The two solutions are related by the interchange of $X^{\pm}$ and
$Y^{\pm}$.
\paragraph{}
Combining vertices of the type described above, we obtain the
diagram shown in Figure \ref{Afig7}a corresponding to the scattering of two
BPS particles with spectral parameters $X_{1}^{\pm}$, $X_{2}^{\pm}$.
If both particles belong to an $SU(2)$ subsector they carry
positive conserved charges $Q_{1}=n(X_{1}^{\pm})$ and
$Q_{2}=n(X^{\pm}_{2})$. The diagram corresponds to the formation of a BPS
boundstate of charge $Q_{1}+Q_{2}$ in the s-channel. These charge
assignments are shown in Figure \ref{Afig7}b.
For individual vertices
the conservation laws can be implemented using
either solution $\alpha$ or $\beta$ described above.
However, once the vertices are connected by an internal line,
consistency requires the choice of the same
solution at both vertices.
If we choose the solution $\beta$ at both
vertices, the spectral parameters of the internal line are fixed to be
$Z^{+}=X_{2}^{+}$, $Z^{-}=X_{1}^{-}$
and the diagram leads to an S-matrix pole when
the parameters of the initial (and final) particles
obey $X_{1}^{+}=X^{-}_{2}$. This pole is present in \nref{bdsfactor}.
In addition, parametrizing the momenta as in \nref{momep} we find that
$q>0$ at this pole. This can be seen most easily by noting that
$u_1 - u_2 =-i/g$ and that $\partial_p u(p) <0$ for real momenta.
 The other possiblity is to choose
the solution $\alpha$ at each vertex which leads to the relation
$X_{1}^{-}=X_{2}^{+}$. This possibility implies that $q<0$ and would not
lead to   bound states. The $S$ matrix may or may not have a pole at this
position, but we should not interpret it as a bound state. For example, in
the $SL(2)$ subsector there is a pole at this position, but we do not have
boundstates associated with them.
\begin{figure}
\centering
\psfrag{t}{\footnotesize{$t$}}
\psfrag{a}{\footnotesize{$a)$}}
\psfrag{b}{\footnotesize{$b)$}}
\psfrag{x}{\footnotesize{$X_{1}^{\pm}$}}
\psfrag{y}{\footnotesize{$X_{2}^{\pm}$}}
\psfrag{v}{\footnotesize{$X_{2}^{\pm}$}}
\psfrag{w}{\footnotesize{$X_{1}^{\pm}$}}
\psfrag{u}{\footnotesize{$Z^{\pm}$}}
\psfrag{Q1}{\footnotesize{$Q_{1}$}}
\psfrag{Q2}{\footnotesize{$Q_{2}$}}
\psfrag{Qp}{\footnotesize{$Q_{1}+Q_{2}$}}
\includegraphics[width=75mm]{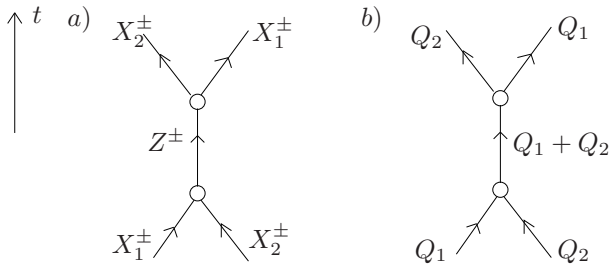}
\caption{Formation of a bound state in the s-channel}
\label{Afig7}
\end{figure}
\paragraph{}
The three-point vertices shown in Figure \ref{Afig6}, are the only possible ones
if all three particles belong to the same $SU(2)$ sector. In any
$SU(2)$ sector, there is a conserved $U(1)$ charge $Q$ such that BPS
particles of type $n$ carry positive charge $Q=n$.
However, the theory also contains BPS particles of negative charge
$Q=-n$ with respect to the same $U(1)$. The particles of negative
charge correspond to a different $SU(2)$ sector.
For each value of $n$, the two particles belong to the same
$SU(2|2)^{2}$ multiplet, but have opposite charges under the
Cartan generator corresponding to the $U(1)$ in question.
The crossing symmetry of the S-matrix suggests that we should also admit
vertices for interactions between BPS particles of positive and
negative charge. For a BPS particle with spectral parameters $X^{\pm}$
the crossing transformation is,
\begin{eqnarray}
X^{+}\rightarrow \tilde{X}^{+}=1/X^{+} & \qquad{} &
X^{-}\rightarrow \tilde{X}^{-}=1/X^{-}
\end{eqnarray}
This transformation changes the sign of the energy and momentum,
$E(\tilde{X}^{\pm})=-E(X^{\pm})$, $p(\tilde{X}^{\pm})=-p(X^{\pm})$ but
preserves the form of the constraint (\ref{const}) with
$n(\tilde{X}^{\pm})=n(X^{\pm})$. If we apply this transformation
to a vertex
with an {\em incoming} state of spectral parameters $X^{\pm}$
and positive charge $Q=n(X^{\pm})$, we obtain a new vertex with an {\em
  outgoing} particle of spectral parameters $\tilde{X}^{\pm}$.
For the new vertex to conserve charge, we must interpret the outgoing
particle as one of negative charge $\tilde{Q}=-n(\tilde{X}^{\pm})=-Q$.
We will indicate a particle of negative charge by reversing direction
of the arrow
appearing on its world line. To illustrate the crossing transformation
we will apply it to an incoming leg on the vertex shown in Figure 2a
to obtain a new vertex as shown in Figure \ref{Afig8}. The solutions $\alpha$
and $\beta$ of the conservation laws for the original vertex yield
two solutions for the new vertex,
\begin{eqnarray}
{\tilde{\mathbf{\alpha}}:} \qquad{} \begin{array}{c} 1/\tilde{X}^{+}=Z^{+} \\
  1/\tilde{X}^{-}=Y^{+} \\ Y^{-}=Z^{-}
\end{array}  & \qquad{} & {\tilde{\mathbf{\beta}}:}
\qquad{} \begin{array}{c} 1/\tilde{X}^{+}=Y^{-} \\
 1/\tilde{X}^{-}=Z^{-} \\ Y^{+}=Z^{+}
\end{array} \label{vertexcrossed}
\end{eqnarray}
A feature of these equations that is worth noting is the
following. If particle $Y^\pm$ is in the giant magnon regime,
% which we denote as type ${\bf G}$,
then one of the two other
particles has to also be in the giant magnon regime while the last one is in the
plane wave regime.
%of type ${\bf G}$ and the other should be
%of type ${\bf P}$ (plane wave regime).
 \begin{figure}
\centering
\psfrag{t}{\footnotesize{$t$}}
\psfrag{x}{\footnotesize{$X^{\pm}$}}
\psfrag{y}{\footnotesize{$Y^{\pm}$}}
\psfrag{z}{\footnotesize{$Z^{\pm}$}}
\psfrag{v}{\footnotesize{$\tilde{X}^{\pm}$}}
\psfrag{w}{\footnotesize{$Z^{\pm}$}}
\psfrag{u}{\footnotesize{$Y^{\pm}$}}
\includegraphics[width=100mm]{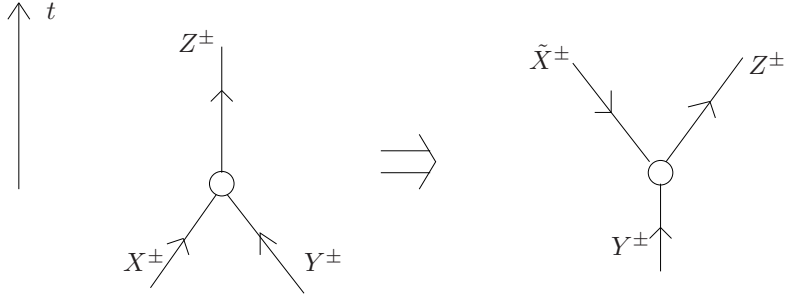}
\caption{Crossing transformation of vertex.}
\label{Afig8}
\end{figure}
\paragraph{}
We can now discuss the pole at $X_1^+ = 1/X_2^-$. This arises naturally if we take the
 s-channel diagram and we cross one of the two particles. We then have a diagram that looks
like the one in Figure \ref{tchannel} by using \nref{vertexcrossed}. Note that the
intermediate particle has $U(1)$ charge zero, but this is not a problem since there are
such particles in the full $SU(2|2)^2$ multiplet corresponding to a BPS magnon with $n=2$.
 \begin{figure}
\centering
\includegraphics[width=50mm]{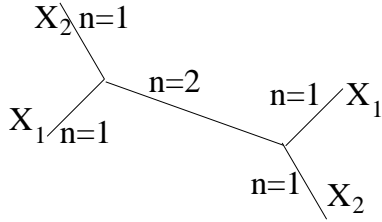}
\caption{t-channel contribution}
\label{tchannel}
\end{figure}
When the external particles are in the near plane wave regime, it is clear that this
is an allowed process since the theory is nearly relativistic.
\paragraph{}
If the external particles are both in the giant magnon region, then, as we mentioned
after \nref{vertexcrossed}, the intermediate particle is in the plane wave regime.
The two giant magnons behave as two heavy particles that are moving at slow relative velocities
which are  interacting through a potential generated by the exchange of the lighter particle.
Let us now remind the reader how these poles arise when we think about the non-relativistic
approximation for the two heavy particles. We will have a lagrangian of the form
$L \sim  M\dot x^2 + e^{ m x } $ where $m x  \ll 0$. When we try to solve the problem in the
Born approximation we are lead to
\be \la{Born}
{ \cal A} \sim \langle \Psi_{out} | V | \Psi_{in}\rangle  = \int dx e^{ - q x } e^{  m x} e^{- q x }
\ee
which gives a divergence when $q = m/2$. Note that this divergence arises from long distance
effects in the quantum mechanics problem and does not depend on the details of the potential
at short distances.
In addition, this feature is independent of the sign of $M$. Indeed, when we expand the
giant magnon dispersion relation we will find that $M$ is negative. The intermediate particle
of mass $m$ ($m>0$) is carrying momentum $k \sim  -i m$. In principle, one could have considered
a situation where this particle carries momentum $k \sim + i m$, but it would have given
rise to an unphysical growing potential. We will use similar criteria in more complicated
situations below in order to isolate the physical singularities from the unphysical ones.
As we mentioned in Section 2, we can have singularities in unphysical sheets from such solutions.

\subsection{Double poles}
%\subsubsection*{The box diagram}
\paragraph{}

\begin{figure}
\centering
\psfrag{t}{\footnotesize{$t$}}
\psfrag{phi}{\footnotesize{$\varphi$}}
\psfrag{psi}{\footnotesize{$\psi$}}
\psfrag{chi}{\footnotesize{$\chi$}}
\psfrag{Q}{\footnotesize{$-Q$}}
\psfrag{Qp}{\footnotesize{$1-Q$}}
\psfrag{Qm}{\footnotesize{$1+Q$}}
\includegraphics[width=70mm]{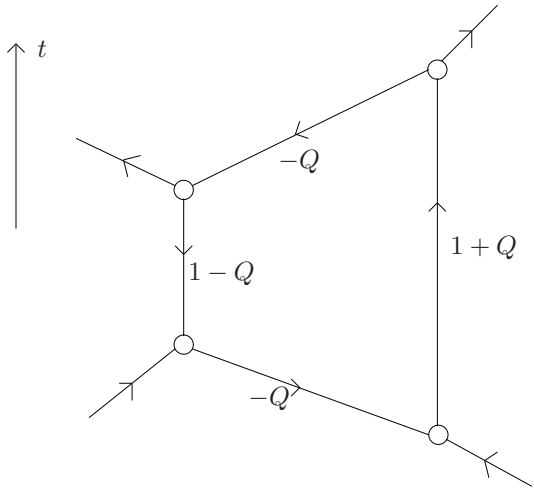}
\caption{The box diagram. $Q>1$.}
\label{Afig4}
\end{figure}

\begin{figure}
\centering
\includegraphics[width=100mm]{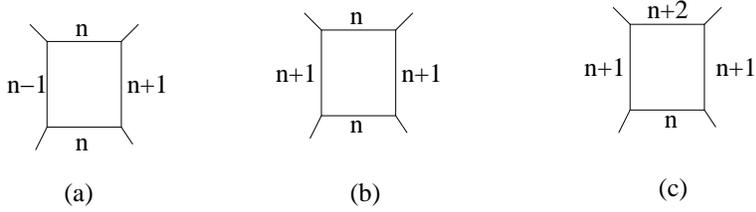}
\caption{Several choices for the $SU(2|2)^{2}$ representations
appearing in the internal lines. In addition we can change the
$\pm 1$ to $\mp 1$ to find new possibilities.} \label{boxdiagrams}
\end{figure}

Following the discussion in Section 2, we look for diagrams that
can give rise to double poles.  Let us start by considering the
one loop box diagram shown in Figure \ref{Afig4}, where the external legs
are all elementary magnons in the $SU(2)$ sector and carry charge
1.
The box diagram\footnote{ This diagram is
different from those considered by Coleman and Thun in the
relativistic sine-Gordon case. In that case this diagram can also
be considered and give singularities that coincide with the ones
found in \cite{ct} using different diagrams. } represents an s-channel
process where the intermediate states consist of two BPS particles of
charges $1+Q$ and $1-Q$ where $Q>1$ is a positive integer. More generally the
states going around the loop can correspond to other members of
the corresponding BPS boundstate magnon multiplet which is
labelled by the positive integer $n$. What will be important for
us is the value of $n$ for each of these magnons since it is the
parameter appearing in the dispersion relation of the state. When
we assign values of $n$ for each of the internal lines we should
remember that, since the external lines have $n=1$, group theory
says that at each vertex the values of $n$ should differ by plus
or minus one.  In figure \ref{boxdiagrams} we have written several
choices. We will first concentrate on the one in figure
\ref{boxdiagrams}a, which is the one suggested by figure
\ref{Afig4}. The choice in \ref{boxdiagrams}c gives the same
condition as the one we will consider explicitly. The choices in
\ref{boxdiagrams}b lead to inconsistent equations once we impose
the condition that the pole is on the physical sheet.
\paragraph{}
To verify this is an allowed process there are two steps,
\paragraph{}
{\bf 1} Implement energy and momentum conservation at each vertex.
\paragraph{}
{\bf 2 } Impose additional
conditions to make sure that the singularity arises
on the physical branch.
\paragraph{}
We will find that after the first step, we will have fixed all the
momenta of the intermediate lines, up to some discrete choices.
The second condition will rule out some of them. In particular, we
can perform approximations when we are evaluating the second
condition, since the exact
position of the poles is already fixed after the first step.
\begin{figure}
\centering
\psfrag{X2pm}{\footnotesize{$X^{\pm}_{2}$}}
\psfrag{X1pm}{\footnotesize{$X^{\pm}_{1}$}}
\psfrag{Y2pm}{\footnotesize{$Y^{\pm}_{2}$}}
\psfrag{Y1pm}{\footnotesize{$Y^{\pm}_{1}$}}
\psfrag{Z1pm}{\footnotesize{$Z^{\pm}_{1}$}}
\psfrag{Z2pm}{\footnotesize{$Z^{\pm}_{2}$}}
\psfrag{t}{\footnotesize{$t$}}
\psfrag{a}{\footnotesize{$A$}}
\psfrag{b}{\footnotesize{$B$}}
\psfrag{c}{\footnotesize{$C$}}
\psfrag{d}{\footnotesize{$D$}}
\includegraphics[width=100mm]{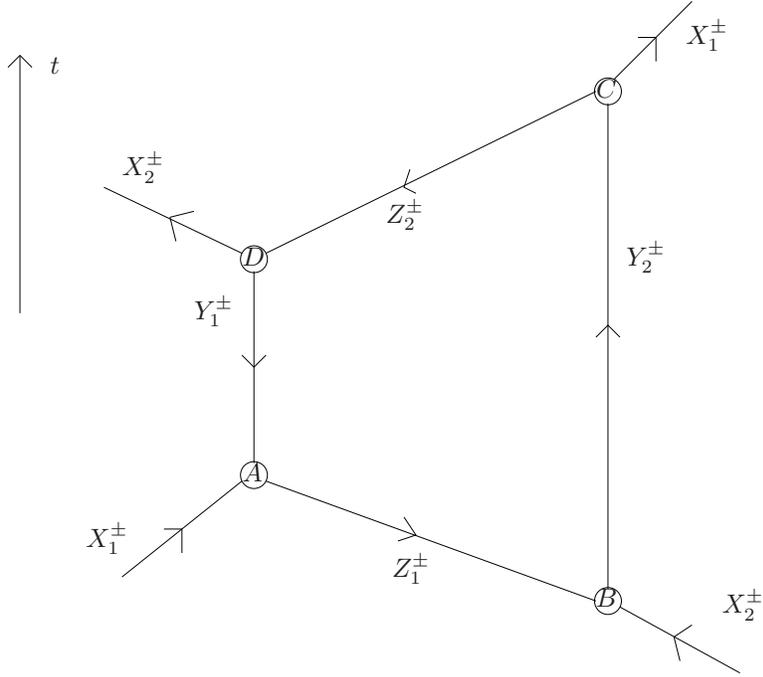}
\caption{Spectral parameters and vertices for the box diagram.}
\label{Afig3}
\end{figure}
\paragraph{}
To perform the first step we assign spectral parameters to each line
in the diagram as shown in Figure \ref{Afig3}. Momentum and energy conservation
are implemented by the vertex rules described
above together with further rules obtained by
repeated applications of the crossing
transformation. There are
two possible choices at each of the four vertices $A$, $B$, $C$ and
$D$. For each of these choices we will then have to evaluate the
second criterion.
\paragraph{}
Let us first consider the case where the two particles are in the
plane wave regime. There we can use the approximate relativistic
formulas and demand that the graphs closes in ``anti-Euclidean'' space.
Each of the vertices involves particles of mass $1$ (the external
line), $Q$ and $Q-1$ (with $Q>1$). The solution to the energy and
momentum conservation condition tells us that the momentum is
approximately zero
so that the energies are given approximately
by $1$ and $Q$ and $Q-1$ respectively. Therefore the
particles $Q$ and $Q-1$ cannot be both be going into the future at
this vertex. It is clear that we cannot obey this condition both at vertices
$A$ and $B$. Thus, none of the graphs gives rise to a singularity in the near
plane wave region.
\paragraph{}
We now consider the case where both incoming particles are
giant magnons. As we mentioned above, one of the particles
emerging from the vertex has to be giant and the other a plane
wave particle. Moreover, the giant particle has energy and
momentum similar to the incoming particle. Thus, we expect that it
continues to move to the future. Since the particle $Z$ is coming
from the past in vertex $A$ or $B$, we conclude that particle $Z$
must be of the plane wave type. Thus, we see that we have the
qualitative conditions $X_1^\pm  \sim Y_1^\pm $ and $Z^+ \sim Z^-
= X_1^\pm$, where the last sign will depend on the type of
solution of the momentum conservation conditions
(\ref{twochoices}) that we choose. For example, for vertex ${\bf A}$ we have
the two choices
\begin{eqnarray}
{\bf A:} \qquad{} \begin{array}{c} X_{1}^{+}=1/Y_{1}^{-} \\
  X_{1}^{-}=1/Z_{1}^{-} \\ Y_{1}^{+}=Z_{1}^{+}
\end{array} & \qquad{} &  {\bf A':} \qquad{} \begin{array}{c} X_{1}^{+}=1/Z_{1}^{+} \\
  X_{1}^{-}=1/Y_{1}^{+} \\ Y_{1}^{-}=Z_{1}^{-}
\end{array}
 \la{aprime}
\end{eqnarray}
Approximately, these to choices amount to saying whether $Z \sim
X_1^+$ (for ${\bf A}$) or $Z\sim X_1^-$ (for ${\bf A'}$).
   For simplicity, let us analyze this condition around
$p=\pi$ (see \nref{momep}). Since this is a maximum in the dispersion relation we can
go to ``anti-Euclidean space'' by setting $p = \pi + i q$.  The
expansion around this point looks similar to the non-relativistic
expansion of a relativistic theory, except for the sign of the
 mass. Moreover, we will have that $q\ll 1$ and thus we will approximate
 $X^+_1 \sim i $, $X_2^- \sim - i $.
  Equating these to $Z$ and parametrizing its momentum as $k = i \kappa$ (see \nref{pp1})
   we find that\footnote{The choice of sign in the square
root corresponds to the sign for the energy.} \bea {\bf A}: ~~~&&
X^+_1 \sim Z_1 ~,~~~~\to i \sim  { n \pm \sqrt{ n^2 - \kappa^2}
\over i \kappa } ~,~~~~ \to \kappa \sim -  n \la{firstf}
\\
{\bf A'}:~~~&& X^-_1 \sim Z_1 ~,~~~~\to - i \sim  { n \pm \sqrt{
n^2 - \kappa^2} \over i \kappa } ~,~~~~ \to \kappa \sim +  n
\la{seconf} \eea
As we mentioned above in our discussion of $t$ channel contributions, we
can interpret the exchange of the $Z$ particle as giving rise to a potential
which will go like $e^{- \kappa x}$. In order to get a sensible potential for
$ x \ll 0$ we must demand that $\kappa <0$. This selects the first condition, ${\bf A}$.

This process can be repeated at each vertex and thus we select only
one particular combination which is,
\begin{eqnarray}
{\bf A:} \qquad{} \begin{array}{c} X_{1}^{-}=1/Z_{1}^{-} \\
  Z_{1}^{+}=Y_{1}^{+} \\ Y_{1}^{-}=1/X_{1}^{+}
\end{array}  & \qquad{} & {\bf B:}
\qquad{} \begin{array}{c} X_{2}^{-}=1/Z_{1}^{+} \\
  X_{2}^{+}=Y_{2}^{+} \\ Y_{2}^{-}= 1/Z^{-}_{1}\end{array} \label{v2} \\
{\bf C:} \qquad{} \begin{array}{c} X_{1}^{+}=1/Z_{2}^{-} \\
  1/Z_{2}^{+}=Y_{2}^{+} \\ Y_{2}^{-}=X_{1}^{-}
\end{array}  & \qquad{} & {\bf D:}
\qquad{} \begin{array}{c} Z_{2}^{+}=1/X_{2}^{+} \\
  Z_{2}^{-}= Y_{1}^{-} \\ 1/X_{2}^{-}= Y^{+}_{1}\end{array} \label{v3}
\end{eqnarray}
or, more simply,
\begin{eqnarray}
 X_{1}^{+}=\frac{1}{Y_{1}^{-}} = \frac{1}{Z_{2}^{-}}  & \qquad{} &
 X_{1}^{-}=Y_{2}^{-}=\frac{1}{Z_{1}^{-}} \nonumber \\
 X_{2}^{+}=Y_{2}^{+}=\frac{1}{Z_{2}^{+}}  & \qquad{} &
\ X_{2}^{-}=\frac{1}{Y_{1}^{+}}=\frac{1}{Z_{1}^{+}}
 \label{equal}
\end{eqnarray}
From these relations as well as the constraints (\ref{const}) on each set of
spectral parameters we derive a condition on the rapidities of the two
incoming particles,
 \begin{equation}
u(X^{\pm}_{1})-u(X_{2}^{\pm})=- \frac{i}{g}\, n ~,~~~~~~~~~ n > 1
\label{pc1}
\end{equation}
for each integer $n>1$. In particular, the graph shown in
Figure \ref{Afig4} gives rise to the double pole with $n=Q$.
In Section 5 we will find that these are
indeed the positions of the poles of the dressing factor $\sigma^{-2}$.
Note that the other ways of solving the energy and momentum
conservation conditions at each vertex, which we have ruled out by the
arguments given above, would have led to equations similar to \nref{pc1} but the integer in the right hand
side would have had a different range.
\paragraph{}
We should note that when we write the condition of the pole as \nref{pc1} we are losing
some information since our analysis leading to \nref{pc1} used that we were in the giant magnon
region. Thus the condition \nref{pc1} only applies in the giant magnon region where the real
part, $r_{i}$ of $u_i $, satisfies $|r_i | < 2$. On the other hand, we have already seen that in the plane wave
region where $|r_i| > 2$ there are no poles. Indeed, we will find that this is a feature of the
proposed S-matrix in \cite{bhl,bes}.

\subsubsection*{Other diagrams}
\begin{figure}
\centering
\psfrag{t}{\footnotesize{$t$}}
\psfrag{phi}{\footnotesize{$\varphi$}}
\psfrag{psi}{\footnotesize{$\psi$}}
\psfrag{chi}{\footnotesize{$\chi$}}
\psfrag{Q}{\footnotesize{$-Q$}}
\psfrag{Qm}{\footnotesize{$Q+1$}}
\includegraphics[width=100mm]{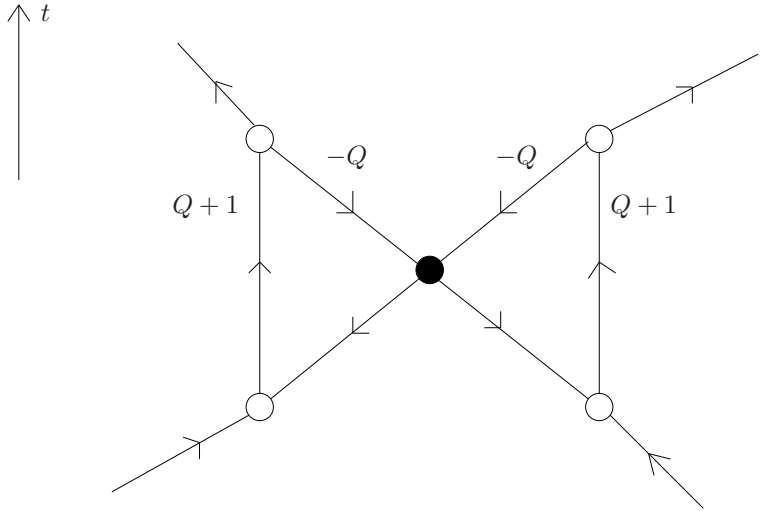}
\caption{The spacetime graph for $t$-channel process.}
\label{Afig5}
\end{figure}

\begin{figure}
\centering
\psfrag{X2pm}{\footnotesize{$X^{\pm}_{2}$}}
\psfrag{X1pm}{\footnotesize{$X^{\pm}_{1}$}}
\psfrag{Y2pm}{\footnotesize{$Y^{\pm}_{2}$}}
\psfrag{Y1pm}{\footnotesize{$Y^{\pm}_{1}$}}
\psfrag{Zpm}{\footnotesize{$Z^{\pm}$}}
\psfrag{t}{\footnotesize{$t$}}
\psfrag{a}{\footnotesize{$A$}}
\psfrag{b}{\footnotesize{$B$}}
\includegraphics[width=100mm]{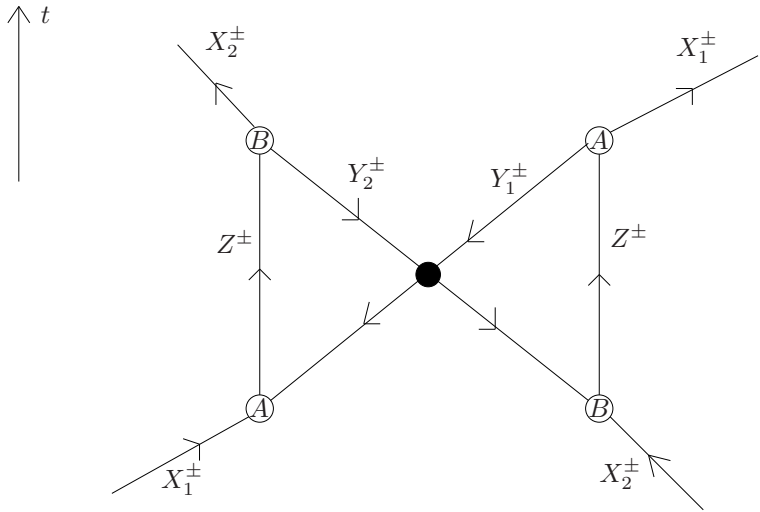}
\caption{Spectral parameters and vertices for the $t$-channel diagram.}
\label{Afig1}
\end{figure}

\paragraph{}
In this section we will analyze other possible on-shell diagrams which
might contribute. In particular we will consider diagrams of the type
originally studied by Coleman and Thun \cite{ct} in the context of the
sine-Gordon model. Since we have already accounted for all the poles in the $S$ matrix,
we expect that these diagrams give the same poles that we have found before.
There are distinct graphs corresponding to s-
and t-channel processes. A candidate
t-channel diagram is shown in Figure \ref{Afig5}. Relative to the process
considered in the previous section, the new feature is that the internal
legs cross in the center of the diagram and we must include the
appropriate S-matrix element in our evaluation of the diagram. In the
cases considered in \cite{ct},
the central scattering process takes place at generic
values of the momenta for which the corresponding S-matrix element is
finite (and non-zero).
In these cases, the extra factor is a harmless phase which
does not affect the analysis of singularities.
This needs to be reconsidered in the present case.
\paragraph{}
To verify the consistency of this diagram we will proceed exactly as
in the previous subsection. First we assign spectral parameters to
each internal line as shown in Figure \ref{Afig1}.
Conservation laws at each
vertex are solved as follows,
%  OLD EQUATION
%\begin{eqnarray}
%{\bf A:} \qquad{} \begin{array}{c} X_{1}^{+}=Z^{+} \\
%  X_{1}^{-}=1/Y_{1}^{+} \\ 1/Y_{1}^{-}=Z^{-}
%\end{array}  & \qquad{} & {\bf B:}
%\qquad{} \begin{array}{c} X_{2}^{+}=1/Y_{2}^{-} \\
%  X_{2}^{-}=Z^{-} \\ 1/Y_{2}^{+}= Z^{+}\end{array}
%\label{v1}
%\end{eqnarray}
\begin{eqnarray}
{\bf A:} \qquad{}
\qquad{} \begin{array}{c} X_{1}^{+}=1/Y_{1}^{-} \\
  X_{1}^{-}=Z^{-} \\ 1/Y_{1}^{+}= Z^{+}\end{array}  & \qquad{} &
  {\bf B:} \begin{array}{c} X_{2}^{+}=Z^{+} \\
  X_{2}^{-}=1/Y_{2}^{+} \\ 1/Y_{2}^{-}=Z^{-}
\end{array}
\label{v1}
\end{eqnarray}
 For the same reasons we discussed above this diagram does not
lead to singularities if the external particles are in the plane wave region. Therefore, let
us assume that the external particles are in the giant magnon region.
As before there are other possible choices for the vertices but only this one is allowed, once
we use the condition that the $Y$ lines are in the plane wave region and that they should
lead to reasonable potentials in the approximate quantum mechanics problem describing the
slow motion of the particles.
The relations (\ref{v1}) lead
to the following condition on the rapidities of the
incoming particles,
\begin{equation}
u_{1}-u_{2}=-\frac{i}{g}\, n ~,~~~~~~~~~~~n > 0
\label{pc10}
\end{equation}
where $n=Q$ is a positive integer. Unlike the process considered in the
previous subsection the case $n=Q=1$ is allowed.
\paragraph{}
Naively the above analysis predicts double poles at the positions
given in equation (\ref{pc10}). These coincide with the double poles we
found earlier from the diagram in Figure \ref{Afig4} except for the case $Q=1$
which appears to predict a new double pole at
$X^{+}_{1}=1/X_{2}^{-}$. However, we still have to consider the
contribution of the ``blob'' in the centre of the diagram shown in Figures
\ref{Afig1} and \ref{Afig5}. In the case $Q=1$ this corresponds to the scattering of two
anti-magnons with spectral parameters $Y^{\pm}_{1}$ and
$Y^{\pm}_{2}$. From the vertex conditions (\ref{v1}) we deduce that
these parameters obey the relation $Y_{1}^{-}=1/Y_{2}^{+}$. This is
clearly a special value of the momentum.
If the  magnon $S$ matrix has a
pole of order $D$ when $X^{+}_{1}=1/X_{2}^{-}$ then, by unitarity,
the S-matrix for two anti-magnons (which is the same as that for two magnons
by crossing on both legs) will
have a zero of degree $D$ when the spectral parameters obey
$Y_{1}^{-}=1/Y_{2}^{+}$. Instead of simply predicting a double pole
via the formula\footnote{As above $N$ is the number of internal lines
and $L$ is the number of loops in the diagram.}
\begin{equation}
D=N-2L=6-4=2,
\end{equation}
the Coleman-Thun analysis then yields a self-consistency equation for
the degree $D$ of the pole,
\begin{equation}
D=N-2L-D=2-D
\end{equation}
with solution $D=1$. This indicates a {\em simple pole} at
$X_{1}^{+}=1/X^{-}_{2}$. Thus this diagram gives also a simple pole. This is
in agreement with the simple pole that we found from the $t$ channel.
It is also in agreement   with the exact
S-matrix where such a pole arises from the BDS contribution to the
$SU(2)$ sector S-matrix. In fact, similar considerations also hold for
higher values of $Q$ where the central scattering corresponds
to a zero of the boundstate S-matrix obtained consistently by
fusion \cite{doreymagthree}.
In these cases the double zero cancels the double pole and the
diagram is finite. Thus, we only obtain the single pole that is given by
the $t$ channel diagram.

\section{Classical, semiclassical and approximate results}

\subsection{Localized classical solutions and the absence  of non-BPS
bound states}
\paragraph{}
In a previous article by two of us \cite{hm}, some localized
oscillating solutions were found. There, it was proposed that these
solutions could  correspond to non-BPS bound states. As we have seen
here, the singularities of the S-matrix are not interpreted as new
particles, but come from physical processes involving the known BPS
particles. In this section we discuss the solutions in \cite{hm} and
explain why they do not give rise to bound states. What happens is
that the ``breather'' solutions in \cite{hm}  can be split into two
BPS magnons with opposite charges. In order to see this, it is not
enough to consider solutions in $R \times S^2$ as it was done
 in \cite{hm}, but one should consider solutions in a bigger subspace
 of $AdS_5 \times S^5$.
 It turns out that it is enough to consider solutions in $R \times S^3$.
Solutions in this subspace were constructed by Spradlin and
Volovich in \cite{sv} (see also \cite{svtwo} and appendix
\ref{svsolutions} ). Their solutions are functions of the
worldsheet coordinates $\sigma^{0,1}$ and of several complex
parameters which we will denote\footnote{They were called
$\lambda_i$ and $\bar \lambda_i$ in \cite{sv}} as $\lambda_i^\pm$.
If we are to ensure that we have a proper real solution we
generically need\footnote{With the exception of the degenerate
cases where the solution collapses to a single BPS soliton of
charge $Q$ (i.e. $\lambda^-_1=\lambda^+_2$).} that the complex
conjugates of the set of parameters $\{ \lambda^+_i \} $ matches
the set of parameters $\{ \lambda_i^- \}$.  The simplest solution
corresponds to a BPS magnon of some charge. This solution is
characterized by two parameters $\lambda^\pm$ which are taken to
be complex conjugates of each other. These parameters can be
identified with the kinematic variables for the BPS magnon as
$X^\pm = \lambda^\pm$ when the conserved $U(1)$ charge $Q$
(denoted $J_{2}$ in \cite{sv}) is positive. When this charge is
negative the correct identification is $X^{+}=1/\lambda^{-}$,
$X^{-}=1/\lambda^{+}$
%The energy and charge for the magnon are
%given by equations $(4.6)$ and $(4.7)$ in \cite{svtwo}.
\paragraph{}
A second, more complicated solution was also considered in
\cite{sv}. This solution depends on four complex parameters
$\lambda_1^\pm, ~ \lambda_2^\pm $. If we identify these parameters
as the parameters of two magnons, $X_i^\pm = \lambda_i^\pm$, the
solution describes the scattering of these two solitons. Also,
if we take \be \label{hmnonbps} \lambda^+_1 = 1/\lambda_1^-~,
~~~\lambda_2^+ = 1/\lambda_2^-~, ~~~~~~(\lambda_1^+ )^* =
\lambda_2^- ~,~~~~~(\lambda_2^+ )^* = \lambda_1^- \ee
 with
a general complex value for $\lambda_1^+$,  we recover the
``breather'' solution of \cite{hm} . Note that the
solution depends on just one complex parameter which corresponds
to the following two real variables: the momentum $p$ of the
breather and the excitation number $\nu$. This solution can be
viewed as coming from the analytic continuation of the scattering
of two zero charge magnons with rapidities $X_i^\pm  =
\lambda_i^\pm$. With this identification $X_i^+ \neq (X_i^-)^*$,
so it seems we can't view 1 and 2 as physical particles.
\paragraph{}
There is, however, a very interesting property of the solutions in
\cite{sv}. They are symmetric  under the exchange of the parameters
$\lambda_1^- \leftrightarrow \lambda_2^-$, without
exchanging\footnote{Equivalently, we can exchange only $\lambda_1^+
\leftrightarrow \lambda_2^+$. }
 the
parameters $\lambda_i^+$. Thus,  we can take the ``breather''
solution we just mentioned and identify the rapidities of two
magnons, instead, as
\begin{eqnarray}
\lambda_{1}^+=X_{1}^{+} & \qquad{} & \lambda_{2}^+=X_{2}^{+} \nonumber \\
\lambda_{1}^-=X_{2}^{-} &\qquad{} & \lambda_{2}^-=X_{1}^{-}
\nonumber
\end{eqnarray}
We now see from \nref{hmnonbps} that $X_i^+ = (X_i^-)^*$. Therefore,
 the same solution can be
viewed as arising from two magnons with kinematic variables
corresponding to real physical variables of two BPS solitons with
opposite charges, both moving with the same velocity. In this way we
clearly see that the ``breather'' solution in \cite{hm} can be
separated into a superposition of two physical
particles\footnote{The sine-Gordon theory also has localized
oscillating solutions which do not correspond to new bound states,
but are superpositions of the ordinary solitons and breathers (see
\cite{Kalbermann:2004fc} for example). }. Therefore, if we we were
to quantize the solution in \cite{hm} we would have a variable
corresponding to the relative position of these two particles and we
would not end up with a bound state.
\paragraph{}
As a side comment, the breather solutions in \cite{hm} for very high
excitation number $n= 2k + c$ (where $c$ is a constant associated
with the energy localized at the boundaries of the excitation in
(\ref{ZWZWZ}) ) look like excitations of the form \be\label{ZWZWZ}
\cdots ZZZW\bar Z \cdots \bar Z \bar W ZZZ\cdots \ee where $W$ and
$\bar W$ are two impurities and there are $k$ $\bar Z$s. This can
split into
 two
  BPS magnon multiplet with charges $n/2$ and $-n/2$, one made of $n/2$ $W$s and the other
  made of $n/2$ $\bar W$s.
\paragraph{}
In \cite{sv}, another solution was considered which was called a
``bound state'' solution. It was constructed by performing an
analytic continuation of the scattering solution for charged
magnons. This solution also has rapidities related as $\lambda_1^+ =
(\lambda_2^-)^*$ and $\lambda_1^- = (\lambda_2^+)^*$. We see that
under the exchange of the two $\lambda_i^-$ we recover the kinematic
variables of two physical magnons. Moreover, these two magnons have
different velocities which implies that this ``bound state''
solution is simply an ordinary (non localized) scattering solution.
It turns out, then, that all possible physical choices for the
parameters $\{\lambda_i^\pm\}$ yield a non-localized scattering
solution\footnote{With the exception, again, of some degenerate
cases given by sets that collapse to the one particle BPS state of
charge Q (i.e. $\lambda_2^{+} =\lambda_2^{-}$).}. Thus, we have concluded that
there is no solution in the literature that looks like a (non-BPS)
bound state. Therefore, there is no reason, from the classical solutions
alone, to think there are more states in the spectrum than the BPS
bound states.
\paragraph{}
This conclusion agrees with our picture from the analysis of poles
in the previous section that already suggested that we should not
be able to construct new non-BPS bound states, at least from the
scattering of elementary magnons.

\subsection{The semiclassical limit of the quantum theory}
\paragraph{}
The processes considered in Section 3, leading to double poles at
positions dictated by (\ref{pc1}), can with be compared directly
with the classical analysis. In the semiclassical limit, $g\gg 1$, the
incoming particles correspond to Giant Magnons of charge\footnote{By
this we mean that the
  charge $Q=1$ carried by these states can be neglected in this
  limit.} $Q=1\sim
0$. The
corresponding classical solution has complex parameters $X^\pm_i = \lambda_i^\pm$ where
$\lambda_i^\pm$ are related as in \nref{hmnonbps}.
%\begin{eqnarray}
%%\lambda_{1}^+=X_{1}^{+} & \qquad{} & \lambda_{2}^+=X_{2}^{+} \nonumber \\
%\lambda_{1}^-=X_{1}^{-} &\qquad{} & \lambda_{2}^-=X_{2}^{-}
%\nonumber
%\end{eqnarray}
On the other hand the intermediate state in the s-channel box diagram
corresponds to two solitons of charges $1-Q\sim -Q$ and $Q+1\sim Q$
with complex parameters,
\begin{eqnarray}
\lambda_{1}^+=1/Y_{1}^{-} & \qquad{} & \lambda_{2}^+=Y_{2}^{+} \nonumber \\
\lambda_{1}^-=1/Y_{1}^{+} & \qquad{} & \lambda^-_{2}=Y_{2}^{-}
\nonumber
\end{eqnarray}
which obey the reality conditions
$\lambda_{1}^+=(\lambda^-_{1})^{*}$ and
$\lambda_{2}^+=(\lambda^-_{2})^{*}$. As the corresponding rapidities
$u(Y^{\pm}_{1})$ and $u(Y^{\pm}_{2})$  vanish and both momenta are
equal to $\pi$, both the constituent solitons are at rest. Note also
that the particles of charge $-Q$, exchanged between the two massive
solitons, have vanishingly small energy in the semiclassical limit.
\paragraph{}
The equalities (\ref{equal}) imply that the classical solutions
corresponding to the incoming and intermediate states are related by
the interchange of parameters,
\begin{equation}
\begin{array}{cccc}
\lambda_{1}^+\rightarrow \lambda_{1}^+, & \qquad{}
\lambda_{2}^+\rightarrow \lambda_{2}^+, & \qquad{}
\lambda^-_{1}\rightarrow \lambda^-_{2}, & \qquad{}
\lambda^-_{2}\rightarrow \lambda^-_{1}
\end{array}
\end{equation}
As this interchange is a symmetry of the two-soliton solution, the
process corresponds to a {\em single} classical field
configuration. This configuration can be interpreted as consisting
of two solitons of opposite charge $\pm Q$ which are both at rest
or as consisting of two solitons of zero charge with complex
momenta $p_{1}=\pi+iq$ and $p_{2}=\pi-iq$. As already established,
there is a   space of such solutions with the sine-Gordon
breather as a special case.

\subsection{Scattering of giant magnons and a non-relativistic
limit}
\paragraph{}
If we consider a single magnon classical solution, as in
\cite{hm}, we find that the solution has fermionic and bosonic
zero modes. These are also called ``collective coordinates''.
 The bosonic zero modes parametrize the space $R \times
S^3$ where $R$ denotes the position of the magnon and $S^3$
parametrizes the orientation of the solution inside $S^5$. In
addition we have fermion zero modes \cite{zerominahan}. The
quantization of all these zero modes should give the whole tower of
BPS magnons. Suppose that we start with a given magnon of momentum
$\bar p$ and consider other  magnons with very similar momentum. We
can describe them by looking at the quantum mechanics of the
collective coordinates. In order to understand what the Hamiltonian
is we can expand the expression of the energy around a reference
momentum\footnote{ In this normalization $k$ is related to
translations in the coordinate $x$ introduced in (\ref{vel}). This
is the normalization where the speed of light in the plane wave
region is one.}, $p = \bar p +  {k \over 2g }$.  Assuming $0< \bar
p<2 \pi$

\def\p2{{p \over2 }}
\def\half{ { 1 \over 2 }}

\begin{eqnarray}
E &=& \sqrt{ n^2 + 16 g^2 \sin^2 \p2 }  =    4 g  \sin \p2
 +
\frac{1}{8 g  \sin \p2  } n^2\nonumber\\
&=&\bar E +   v   k - { 1 \over 8 g  }  |\sin { {\bar p
\over2 } }|   k^2+ { 1 \over  8 g | \sin { \bar p \over 2 }|
} n^2 \la{energcla}
\end{eqnarray}
\noindent with
\be
  v = { d E \over d k} = \cos { \bar p \over 2}  ~~~~~~~
 \bar E = 4 g | \sin {\bar p \over 2 }|
\ee
\paragraph{}
As we saw above,   diagrams giving rise to
double poles involve ``heavy'' giant magnon particles
 and ``light'' plane wave particles.
  Double poles arise when two heavy particles exchange a pair
of light particles. In this regime, where the relative motion of the
heavy particles is slow, we expect that we should be able to view
the problem as  non-relativistic motion of the collective
coordinates plus a potential that arises from the exchange of the
light particles. Since the dispersion relation for the light
particles is approximately relativistic the potential will have the
usual Yukawa form for particles of mass $m$, i.e. $V \sim e^{ - m r}
$ for $ m r \gg 1$. However, we should recall that the two magnons
are moving with similar velocities which are large, $v_1 \sim v_2
\sim v$. If we denote by $x$ the distance between the two magnons,
we find that the potential has the boosted form \be
 V \sim \sum_n
 g  f_n(\Omega_1,\Omega_2) e^{ - n |x| \gamma } ~,~~~~~~~\gamma^{-2} =
1 - v^2 \ee where $f_n$ is a function that depends on the angles on
both $S^3$s and the fermion zero modes. We have also indicated how
the potential scales with $g$. This scaling is determined as
follows. We know that the classical solutions are independent of
$g$. Thus, classical scattering properties, such as time delays are
independent of $g$. Since the mass is proportional to $g$ (see
\nref{energcla}), then, the potential should also be proportional to
$g$ such that the classical equations of motion are independent of
$g$.
\paragraph{}
It is now convenient to rescale $x \to \hat x = \gamma x$ and $k \to \hat k = k/\gamma$. We can
also remove the linear term in the momentum in \nref{energcla} by defining a new energy
$\hat \epsilon$ which is a linear combination of the old energy and the momentum, $\hat \epsilon
= \epsilon - v k $.
In these new variables we have a Hamiltonian

\begin{equation}\label{H12}
\hat H  = { 1 \over 8 g  |\sin { \bar p \over 2 }| } \left[ -\hat k_1^2 -\hat
k_2^2 + {n}_1^2 + {n}_2^2 + \sum_n g^2  f_n(\Omega_1,\Omega_2)
e^{ - n |\hat x_1 - \hat x_2|} \right]
\end{equation}
\paragraph{}
This is the form for the quantum mechanics of the collective coordinates in
the slow relative velocity regime. We expect that the potential should
be determined completely by
the symmetries plus integrability, but we will not work it out here.
\paragraph{}
It is worth mentioning that, in these variables, the position of
the double poles (see section 3) at $u_1 - u_2 =  - in/g$ is \be
\hat k = \half ( \hat k_1 - \hat k_2 ) =   i n ,~~~~~{\rm with }
~~ n = 2,3,4,... \ee where $\hat k$ is the relative momentum
conjugate to $\hat x_1 - \hat x_2$.

\subsection{A toy model for double poles}
\paragraph{}
Since finding the full potential in \nref{H12} is beyond the scope
of this paper we will consider a toy model which we can solve and is
such that it also displays the double poles. Notice that in a
non-relativistic quantum mechanics problem with one degree of
freedom the scattering matrix can have only single poles
\cite{lamb}. In fact, in the quantum mechanical problem the poles
can have different origins. We can have poles corresponding to bound
states and we can have poles that arise because we have a potential
decaying like $e^{ - m r}$ and there is a resonance between this
exponential decay and the increase of the wavefunction as we discussed
around \nref{Born}.
  It should be emphasized
that the pole originates from the tail of the potential and is
independent of the details of the potential for small $r$. In the
full relativistic theory, the bound states give rise to poles in the
$s$ channel and the tails of the potential
 to poles in the $t$ channel.
\paragraph{}
The double poles we are considering also arise from the tails of the
potential and are not related to bound states. In order to see them,
we should include at least two degrees of freedom. We can include an
internal angular coordinate $\varphi$ which lives on a circle. If we
have a potential of the form $V \sim e^{ in \varphi} e^{ - m_n |x| }
$, then we will have double poles which can be interpreted, in a
relativistic context, as coming from the exchange of a pair of
massive particles.
\paragraph{}
In our  toy model,
 instead of the full space of magnon collective coordinates, we will have $R \times S^1$.
We denote by $\varphi = \varphi_1 - \varphi_2$ the relative angular
coordinate. We consider a theory where we only treat the relative
coordinates $x$ and $\varphi$. We now look for an integrable
potential which contains the exchanged particles charged under the
$U(1)$ symmetry that shifts $\varphi_i$. We recall that the
non-relativistic limit considered here has a counterpart in the
sine-Gordon theory \cite{Zam}. In that context, integrable
potentials of the form $V \sim \frac{1}{\sinh^2 x}$ arise.
 With this in mind we can consider a simple
generalization of this case that will fit our purposes. We propose
to study the following hamiltonian
\begin{equation}\label{Hint}
H = { 1 \over 2} \left[ - \hat k^2 +\ell^2\right] +
\frac{ v}{4 \sinh^2 \left(\frac{\hat x + i  \varphi}{2}\right)} +
\frac{  v }{4 \sinh^2 \left(\frac{\hat x - i
\varphi}{2}\right)} ~,~~~~~~~v = { 1 \over 4} - (m + \half)^2
\end{equation}
\noindent where $v$ is a coupling constant (which we have
parameterized in terms of another coupling $m$), $\hat k$ is the
momentum conjugate to the relative coordinate and $\ell$ is
conjugate to $\varphi = \varphi_1 - \varphi_2$.
 As advertised,
this model is integrable (for reviews on these Calogero-Sutherland
systems see \cite{Olshanetsky:1981dk,Olshanetsky:1983wh,Polychronakos:2006nz}),
 has the right classical limit
and coupling dependence, as long as we make $v \sim
\mathcal{O}(g^2)$. This system is naturally defined on the
cylinder $R\times S^1$. Since we are considering the scattering of
identical particles the space is really $(R \times
S^1)/\mathbb{Z}_2$.
\paragraph{}
This potential is singular at $z=0$ and  near the singularity it
behaves as
  $\cos(2 \theta) /r^2$ where $z =r e^{i \theta}$. Unfortunately,
since the coupling constant is large, this means that
the potential is attractive for some angles, and particles might ``fall to the center''.
This can be avoided by modifying the potential near the singularities. Since the poles
we are interested in
come from long distances, this should have no effect. We do not
know of an easy way to modify the potential while preserving integrability. However,
for particular values of the coupling $v$ there exist explicit solutions of the above
problem where the wavefunction vanishes at the singularities. Thus, a  modification
of the potential at the singularities to make the problem well defined  should
have no effect on these particular wave functions.
\paragraph{}
In order to solve this problem we
 define the following quantities
\begin{eqnarray}
z = \frac{1}{2} (x + i \varphi)  &\quad& \bar z = \frac{1}{2} (x - i \varphi)\\
\beta^2 +\bar \beta^2 = 4 \hat \epsilon \quad &\textrm{with}& \quad
H=\hat \epsilon = \frac{\ell^2}{2} - \frac{\hat k^2}{2}\label{betae}
%\\
% v= 4 q &\quad& \bar v = 4 \bar q
\end{eqnarray}
Therefore the Schrodinger equation we need to solve is
\begin{equation} \label{zprob}
\left[\frac{\partial^2}{\partial z^2} - \beta^2 + \frac{v}{\sinh^2
z} \right] \psi = 0  ~,~~~~~~~v = { 1 \over 4} - (m + \half)^2
\end{equation}
\noindent and its barred version.   Then, the general form of the
solutions to this equation (and its bared version) are
\begin{equation}
\psi^\beta (z) = e^{\beta z} F(1+m,-m,1-\beta,u) \quad
\textrm{with} \quad u =\frac{1}{2}\left(1-{ 1 \over \tanh(z)}\right)
\end{equation}
\noindent where $F$ is the ordinary hypergeometric function $_2F_1$.
The second independent solution is $\psi^{-\beta}(z)$. Therefore,
the general form of the solutions to our whole problem is
$\Psi^{\beta,\bar \beta}(z,\bar z) = \psi^\beta(z)\psi^{\bar
\beta}(\bar z)$ with $\beta$, $\bar \beta$ satisfying (\ref{betae}).
We can check that in the large $x$ region the wave function behaves
as
\begin{equation}\label{fout}
\Psi^{\beta,\bar \beta}(z,\bar z) \sim e^{ \beta z + \bar \beta \bar z} \sim e^{i \hat k x + i \ell
\varphi} ~,~~~~~~x\gg 0 ~,~~~~~~~~\beta = \ell + i \hat k ~,~~~~\bar \beta = -\ell + i
\hat k
\end{equation}
where $\ell$ is integer, since $\varphi$ is periodic. One important
condition is that the wave function remains normalizable in the
neighborhood of the singularity located at $z = \bar z =0$. In fact,
we see that two independent solutions of \nref{zprob} go like
$z^{-m}$ and $z^{1+m}$. Let us assume that $m$ in integer \footnote{
Curiously, for integer $m$, these models naturally arise from matrix
quantum mechanics \cite{Polychronakos:2006nz}. At these values of
the coupling the potential $1/\cosh^2 z$ is also reflectionless. We
also see that  \nref{fout} becomes an order $m$ polynomial of $u$.}.
 We then  see that if we demand that
the wavefunction is antisymmetric\footnote{We actually need
wavefunctions that have definite parity. It turns out that
symmetric ones have a singular behavior at the singularity, so we
need to consider antisymmetric ones. Actually, there is a physical
interpretation of why only antisymmetric functions survive. As the
interaction has a $\cos 2 \theta$ term near the origin, only odd
angular momentum states are such that the potential averages to
zero over an orbit. Symmetric states, thus, fall to the center.}
under $z \to -z$ and $\bar z \to - \bar z$ we need to combine
solutions so that we have combinations such as $z^{-m} {\bar z}^{1
+ m}$ and $z^{1 + m} {\bar z }^{-m}$ near the origin. We see that
such combinations vanish as we approach $ z \to 0$ so that a small
modification of the potential is not expected to affect the
solutions in an important way. In terms of the functions
\nref{fout} we need to consider the following combination
\begin{equation}\label{wave}
\Psi(z,\bar z) = - \Psi(-z,-\bar z) =  \frac{\Gamma(1+m-\beta)\Gamma(1+m-\bar
\beta)}{\Gamma(1-\beta)\Gamma(1-\bar \beta)} \Psi^{\beta,\bar
\beta} -\frac{\Gamma(1+m+\beta)\Gamma(1+m+\bar
\beta)}{\Gamma(1+\beta)\Gamma(1+\bar \beta)} \Psi^{-\beta,-\bar
\beta}
\end{equation}
The fact that the wavefunction is antisymmetric suggests that it is a natural solution
to the problem of scattering of identical fermions, rather than bosons.
  Using the
asymptotic expression (\ref{fout}), we can read off the S-matrix
of this problem from the coefficients in (\ref{wave}) to be
\begin{equation}
S(\beta,\bar \beta) = - \frac{\Gamma(1+m -\beta)\Gamma(1+m-\bar
\beta)}{\Gamma(1-\beta)\Gamma(1-\bar
\beta)}\frac{\Gamma(1+\beta)\Gamma(1+\bar
\beta)}{\Gamma(1+m+\beta)\Gamma(1+m+\bar \beta)}
%\quad \textrm{for} \quad p>0
\nonumber\\
\end{equation}
If we scatter particles 1
and 2 with equal angular momentum $J_2$, then the relative angular momentum vanishes,
$\ell =0$ and  $\beta=\bar \beta = i \hat
k$. In this situation the S-matrix is
\begin{equation}
S = -\frac{( 1- i \hat k )^2( 2- i \hat k )^2( 3- i \hat k )^2...(
m- i \hat k )^2}{( 1+ i \hat k )^2( 2+ i \hat k )^2( 3+ i \hat k
)^2...( m+ i \hat k )^2}
\end{equation}
\paragraph{}
This expression has double poles at $\hat k = i n$ with
$n=1,2,3,...,m$. We discussed in the last subsection that, from
dynamical arguments, the coupling of the quantum mechanical model
for the collective coordinates needs to be of order
$\mathcal{O}(g^2)$. Therefore we need $m \sim \mathcal{O}(g)$. In
the large $g$ regime  we obtain the infinite series of double poles
that we expected from the complete theory. However, we also obtain
an extra double pole at $\hat k =  i$. This is not unexpected as
our toy model does not forbid $J_2=0$ states for the heavy
particles\footnote{Note, however, that they are indeed excluded for
the intermediate ``light'' states. This can be seen by expanding the
effective potential at large distances.}.
Note that there is no single pole related to the fact that, in the toy model,
 the exchanged particles
change $\ell$. In the true theory there is a single pole at $ \hat k = i $.
One the other hand if we look at
the
S-matrix for $\ell \neq 0$ we find a simple pole at $\hat k =
 i |\ell|$ representing the t-channel exchange of the particle that is giving
 rise to the potential in the toy model. In this case, we also
  retain double poles for $\hat k =  i r$ with
$r=1+|\ell|,2+|\ell|,...$
\paragraph{}
 We conclude that this very simple model presents many of the
characteristics associated with the complete S-matrix of the full
string theory. Of course, it would be nice to
find the correct quantum mechanics theory
that describes the full problem in this regime.

\section{Poles in the Beisert-Eden-Hernandez-Lopez-Staudacher S-matrix}

\subsection{Integral expression for the dressing factor}
\paragraph{}
In \cite{bhl,bes} a conjecture was made for the exact form of the unknown
dressing factor that appears in the magnon S-matrix. This dressing
factor was expressed as \be \la{dressf}
%\eqalign{
\sigma^2 =
% &
{ R^2(x_1^+,x_2^+) R^2(x_1^-,x_2^-) \over R^2(x_1^+,x_2^-)
R^2(x_1^-,x_2^+) }
%\cr
~,~~~~~~~~~~R^2(x_1,x_2) =
%&
 e^{ i 2 \chi(x_1,x_2) - i 2 \chi(x_2,x_1) }
% }
\ee
\paragraph{}
The function $\chi$ was given as a series expansion in powers
of $1/x_i$. In order to study its analytic structure it is more
convenient to write it as
 an integral expression. We find it  convenient to introduce
 a new function $\tilde \chi(1,2) $, which differs from $\chi(1,2)$ in \bes\ by terms that
 are symmetric under $1 \leftrightarrow 2$. Such symmetric terms cancel in \nref{dressf} and thus $\tilde
 \chi$ will lead to the same dressing factor if we use it in  \nref{dressf}.
In appendix A we derive  the following integral expression for
$\tilde \chi$ from the formulae in \bes\
 \be \la{integrexp} \tilde
\chi(1,2) =-i \oint {dz_1 \over 2 \pi } \oint { dz_2 \over 2 \pi }
 { 1
 \over
(x_1 - z_1) ( x_2 - z_2) }   \times \log\Gamma( 1 + i   g  ( z_1 +
{ 1 \over z_1}  - z_2 - { 1 \over z_2} ) ) \ee where the integral
is over the contours with $|z_1|=|z_2|=1$. With these contours the
integral is well defined and there are no singularities on the
integration contour if we assume that $|x_1| , |x_2| > 1$, as it
is the case for physical particles.

\subsection{Poles of the dressing factor in the giant magnon region}
\paragraph{}
As we start analytically continuing in $x_1$ and $x_2$ we might
encounter singularities. Singularities will appear when poles or
branch cuts of the integrand pinch the contour. We will study only
a subset of all possible singularities. We will set $g$ to be
large and then we will focus on the giant magnon region with fixed
$p$, there $x^\pm$ are near the unit circle but away from $x^\pm
=1$. We will then analytically continue away from physical values
of $x^\pm$ in this region, but we will still stay near the unit
circle and far from $x^\pm =1$. In other words, starting from the
original values for $p_1$ and $p_2$ we will analytically continue
only in a small neighborhood of these values of order $1/g$ around
the physical values.
\paragraph{}
In order to find the poles of  $\tilde \chi$ in \nref{integrexp},
we view it as a function of three independent variables: $g$,
$x_1$ and $x_2$. In order to avoid having to keep track of the
branch of the log in \nref{integrexp} we take a derivative of
$\tilde \chi$ with respect to $g$. This will allow us to isolate
the singularities of $\tilde \chi$ that depend on $g$. In fact,
there are no singularities that are independent of $g$ because we
can set $g=0$ and we see that $\tilde \chi$ is identically zero.
Thus, taking the derivative and disregarding some terms that do
not contribute to the integral we find \bea  g^2
\partial_g \tilde \chi &=& \sum_{n=1}^\infty   g^2 \partial_g
\tilde \chi_n  \notag
\\
g^2 \partial_g \tilde \chi_n &= & -  \oint { dz_1 \over 2 \pi } {
d z_2 \over 2 \pi }
 { 1
 \over
(x_1 - z_1) ( x_2 - z_2) }     \times {n \over
 - i { n \over g } + z_1 + { 1 \over z_1} - z_2 - {1 \over z_2}  }
 \la{chisnter}
 \eea
We now do the integral over $z_1$ by deforming the contour,  which
starts out at $|z_1|=1$,
 towards the origin
of the $z_1$ plane. The only poles we find come from the last
factor in \nref{chisnter}. The pole is at $z_1 = z_n(z_2)$ were
the function $z_n(z)$ is defined as the solution to \be
\la{solequ} { z_n} + { 1 \over z_n} - z - { 1 \over z } = i { n
\over g } ~,~~~~~~~~~n>0 \ee as in \bhl . Out of the two roots of
this equation we should pick out the root whose absolute value is
 less than one.
We can select this root unambiguously if $|z|=1$. So from now on
we only talk about this root and its analytic continuation as a
function of $z$. We thus find that the result of the $z_1$
integral is
\be
 \la{resinte} g^2 \partial_g \tilde \chi^n =
 - i \oint { dz  \over 2 \pi }    {
 1\over
(  x_2 -z ) (x_1 - z_n(z) ) }    \times {n \over
 1 - { 1 \over z_n^2(z) }  }
 \ee
where we have set $z =z_2$.
\paragraph{}
We would like to understand where \nref{resinte}  has poles when
we change $x_i$. Note that if $|x_i|>1$ (at it is the case for
physical values) then the integral is finite. As we start
analytically continuing $x_i$ we could have a situation where the
pole at $z_2 =  x_2$ moves inside the unit circle.  In that case,
we might think that all we need to do is to deform the contour so
that $|z_2| <1$, but this could push $z_n(z_2)$ to larger values
in such a way that it becomes equal to $x_1$. In other words,
there is pole only if two poles of the integrand pinch the
integration contour. We discuss below when this happens. If the
contour is indeed  pinched then the integral is equal to $- 2 \pi
i $ times
 the residue
at $z_2 = x_2$. This gives \bea \notag g^2 \partial_g \tilde
\chi^n &=& - { 1 \over x_1 - x_n(x_2) } { n \over 1 - { 1 \over
x_n^2 (x_2 )}}
\\ \la{residuein}
\tilde \chi &= & -i \log ( x_1 - x_n(x_2) ) \eea We then find that
$ e^{ 2 i \tilde \chi} \sim   (x_1 - x_n(x_2) )^2 $ has double
zeros.

\subsection{When do we pinch the contour?}

%\ifig\znplot{ Plot of $z_n(z)$ and $1/z_n(z)$ for $z$ on the unit
%circle. The circle gets mapped to an interval. The $z_n$ is inside
%the unit circle.  We have set $n/g = 0.1$. }
%{\epsfxsize1.5in\epsfysize1.25in\epsfbox{znplot.eps}}

\begin{figure}
\centering \includegraphics[width=40mm,height=33mm]{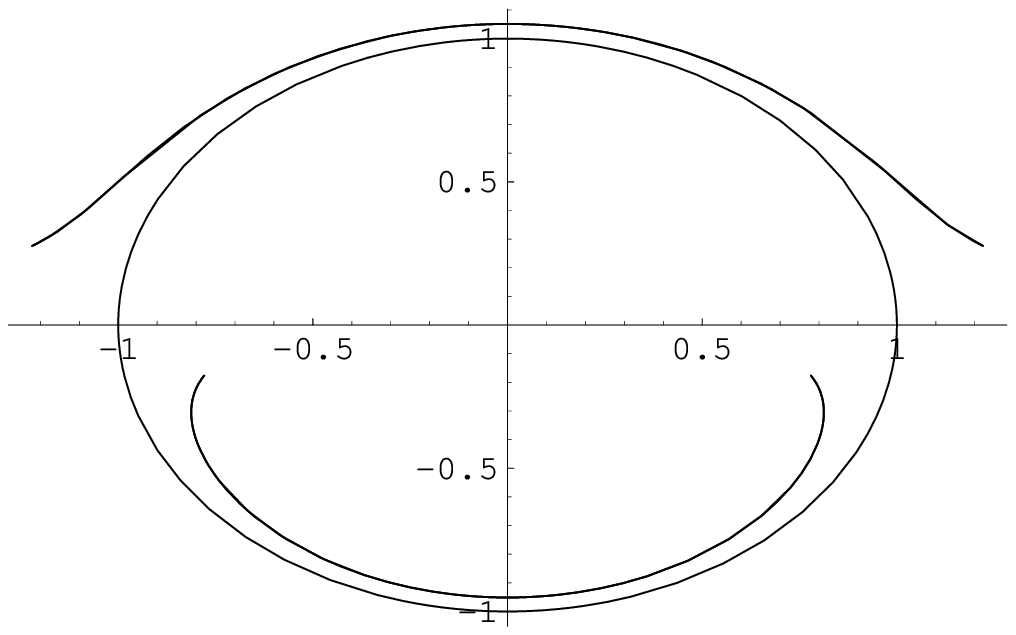} \,
\, \, \, \, \, \, \, \,
\includegraphics[width=40mm,height=33mm]{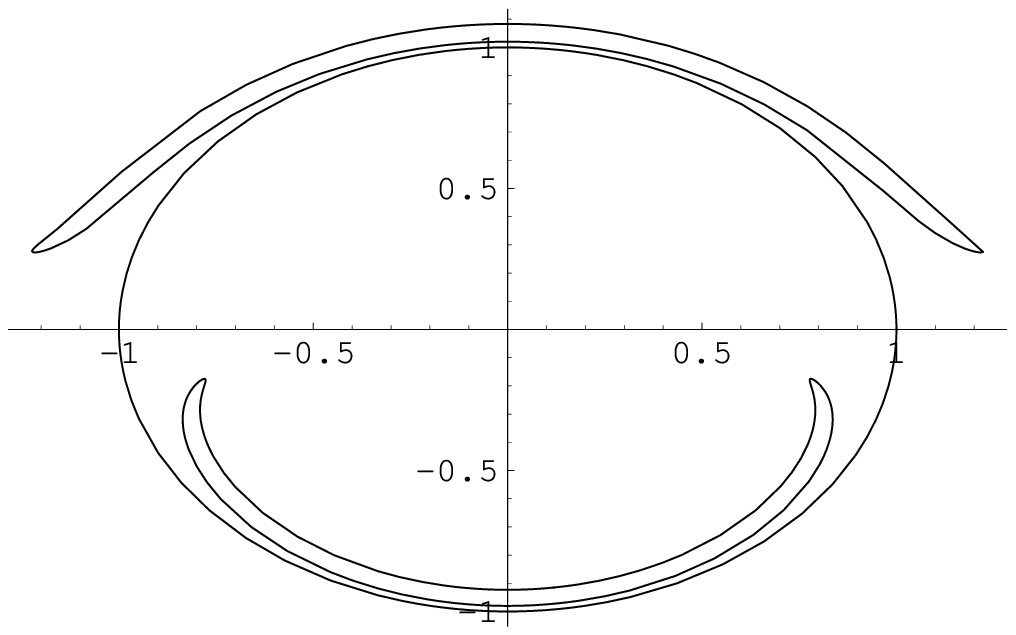}
\caption{On the left we see plots of $z_n(z)$ and $1/z_n(z)$ for
$z$ on the unit circle. The circle gets mapped to an interval.
$z_n$ is the one inside the unit circle. We have set $n/g = 0.1$.
In the plot on the right we have taken $|z| = .98 <1 $. }
\label{znplot}
\end{figure}

%\ifig\znplottwo{ Plot of $z_n(z)$ and $1/z_n(z)$ for $z= e^{-.02}
%e^{i \theta}$.
% Again, $n/g = 0.1$. }
%{\epsfxsize1.5in\epsfysize1.25in\epsfbox{znplottwo.eps}}
\paragraph{}
We need to understand some aspects of the function $z_n$ more
precisely (see Figure \ref{znplot}). The essential feature is the
following.
 If we set $z = e^{i \theta}$ then $z_n \sim 1/z$ for $ 0< \theta< \pi $ and
$z_n \sim z$ for $ -\pi < \theta < 0 $. Recall also that in the
giant magnon region the physical values of $x^\pm \sim e^{ \pm i
p/2}$. Both are close to the unit circle but $x^+$ is in the upper
half plane and $x^-$ in the lower half plane.

%\begin{figure}
%\centering \includegraphics[width=40mm,height=33mm]{znplottwo.eps}
%\caption{ Plot of $z_n(z)$ and $1/z_n(z)$ for $z= e^{-.02} e^{i
%\theta}$.
% Again, $n/g = 0.1$.  } \label{znplottwo}
%\end{figure}
\paragraph{}
Suppose that we have $x_2 = x_2^+$. This will be on the upper half
plane and for finite and physical (real) values of $p_2$ it will
be close to the unit circle, but outside the unit circle. If we
keep $x_1$ fixed and we analytically continue in $x_2^+$ then we
see that $z_n$ will be in the lower half plane when $z \sim
x_2^+$, thus we can only pinch the contour if $x_1$ is also in the
lower half plane. This happens when  $x_1=x_1^-$, but not for $x_1
= x_1^+$. Since $z_n \sim 1/z$ for $z \sim x_2^+$ we find that
 $|z_n|$ increases as $|z|$ decreases (see Figure \ref{znplot}).
Thus we pinch the contour when $x_2 = x_2^+$ and $x_1 = x_1^-$.
\paragraph{}
Now suppose that $x_2 = x_2^-$.  If $z \sim x_2^-$ we have  $z_n
\sim z $ so that  when we decrease $|z|$ we will also decrease
$|z_n|$ and we will not pinch the contour.
So the only case where we pinch the contour is when $x_2^+  = z$
and $x_1^- = z_n$. Of course, we have discussed poles from $\tilde
\chi(1,2)$. When we consider $\tilde \chi(2,1)$ we have the same
story with
 $1 \leftrightarrow 2 $.
\paragraph{}
Thus, the final result  is that we have poles and zeros of the
form \be \la{poles} \sigma^2 \sim  e^{ - 2 i ( \tilde
\chi(x_1^-,x_2^+) - \tilde \chi(x_2^-,x_1^+) ) } \sim \prod_{n>0}
{  (x_2^- - x_n(x_1^+) )^2 \over (x_1^- - x_n(x_2^+) )^2 } \ee We
have only indicated terms that give rise to poles or zeros in the
region of interest (large $g$ , $p_1 \sim p_2$  of order one and
$p_1-p_2 \sim \mathcal{O}(g)$). Of course, the expression for the
location of poles is exact and we will find poles at these
locations in the appropriate branch for any values of $p_i$ and
$g$. \footnote{ Note that there are not poles or zeros at $x_1^- =
x_n(x_2^-) $ in the branch describing the neighborhood of the
giant magnon region. Such poles are probably present on another
branch. }
\paragraph{}
The double zeros of $\sigma^2$ (or double poles of $\sigma^{-2}$)
 lie at \be \la{finpos} x_2^- = x_n(x_1^+) ~,~~~~~~n>0
\ee This implies that \be \la{newimp} u_1 - u_2  = - {  i (n+1)
\over g } = - i { m \over g} , ~~~m > 1 \ee Thus we see that the
poles do not start at one. In fact, we do not get poles   in
$\sigma^{-2} $ for $m=1$.\footnote{ Notice, however, that there
are poles (or zeros) at that position in the one loop factor
answer in \bhl\ (see also \cite{hl}).
 These poles (or zeros) are cancelled by all the higher order contributions.}
Note that equation \nref{finpos} contains more information than
\nref{newimp} since we have specified a particular branch of the
function $x_n$ . In fact, if we only looked at \nref{newimp} we
might incorrectly conclude that there are poles in the near plane
wave region with $x_1^\pm  \sim x_2^\pm$ and $|x_i^\pm| \gg 1 $.
On the other hand, we see from \nref{finpos} that in this region
$x_n(x_1^+) \sim 1/x_1^+$, whose absolute value is not much larger
than one. Of course, it is obvious from the integral expression
\nref{integrexp} that there are no poles in the region $|x_i| > 1$
since the integral is explicitly finite there.
\paragraph{}
In this whole discussion we have assumed that $n \ll g$. If this
were not the case, we would have to move by a larger amount from
the giant magnon region and we would have to understand better the
whole analytic structure of the function. Finally we note that a
related integral representation of the BHL/BES phase has recently
been obtained in \cite{kostov}. Like Eqn (\ref{chisnter}) above,
the formula (5.11) appearing in this reference is also suggestive
of a sum over intermediate states involving BPS particles.

\section{Conclusions}
\paragraph{}
We have understood the physical origin of the poles in the S
matrix for the scattering of fundamental magnons. These poles can
be cleanly isolated for strong coupling in the giant magnon
region. In this case, the poles are far from other singularities of the S
matrix. These poles can be explained by the interchange of BPS
magnons. The origin of the double poles in the magnon S-matrix is
the same as the origin of the double poles in the sine Gordon
S-matrix. The position of the poles is determined by  the
spectrum of BPS particles of the model.
 These poles are accounted for by considering all the BPS particles
that were found to exist on this chain.
 Therefore there is no
reason to think that the spin chain has any other bound states
beyond the BPS ones we already know about. In fact, the localized
non-BPS classical solutions that were found in \hm\ were shown to
be continuously connected to separated BPS magnon configurations.
Thus, those solutions do not correspond to new states.
One thing that we have not explored is the physical origin of the
branch cuts in the dressing factor. It would be very nice to
understand these more precisely.
% {\bf Acknowledgments }
\paragraph{}
ND is supported by a PPARC Senior Fellowship.
JM was supported in part by DOE grant \#DE-FG02-90ER40542.

\appendix

\section{Derivation of \nref{integrexp} }

We start from the expression for $\chi(x_1,x_2)$ as in equation
(A.9) of   \bes \bea \la{chisum}
 \chi &= & - 2 \sum_{r=2}^\infty \sum_{s =
r+1}^\infty \cos( { \pi \over2 } (s -r-1) ) { 1 \over x_1^{r-1}
x_2^{s-1} } \int_0^\infty {d t \over t} { J_{r-1}(2 g t) J_{s-1}(
2 g t) \over e^t -1 } \\
\notag \chi &= & - 2 { 1 \over x_1 x_2^2 }
\sum_{w=0}^\infty\sum_{l=0}^\infty (-1)^l  { 1 \over (x_1 x_2)^w
x_2^{ 2l } } \int_0^\infty {d t \over t} { J_{w +1}(2 g t) J_{2 +
2 l }( 2 g t) \over e^t -1 }
\eea
 where $s = r+1 + 2 l $ and $r = w
+ 2$. We then use the following expression for the Bessel
functions
\be \la{frmj} J_n (z) =  \int_0^{2 \pi } { d \theta \over2
\pi } e^{ - i n \theta + i z \sin \theta  }
\ee
 We insert these
expressions in the above formulas, we perform the sums and after a
simple shift of the integration variables we obtain
\bea \la{chisumtw}
\chi &= & 2 i  { 1 \over x_1 x_2^2 }
\int_0^{2 \pi } { d \theta_1 \over2 \pi }
 \int_0^{2 \pi } { d \theta_2 \over2 \pi }   e^{ -i \theta_1 - 2 i \theta_2} {
 1\over
(  1 -  {  e^{ - 2 i \theta_2 } \over x_2^2 } ) (1 - { e ^{ - i
\theta_1 - i \theta_2 } \over x_1 x_2 } ) } \times \\ \notag && ~~~~
\times \int_0^\infty {d t \over t } {  e^{ i ( 2 g t ) ( - \cos
\theta_1 + \cos \theta_2 )} \over e^t -1}
\eea
 We now use that
\be \la{zetaid}
 \int_0^\infty { d t \over t } { e^{ - z t} \over e^t
-1 }  = C_1 + z C_2 + \log \Gamma(1 + z)
\ee
 where $C_1$ and $C_2$
are divergent constants which do  not contribute once we do the
integral over $\theta_i$. We then find
\bea \la{chitw}
 \chi &  =
& 2 i  { 1 \over x_1 x_2^2 }  \int_0^{2 \pi } { d \theta_1 \over2
\pi }
 \int_0^{2 \pi } { d \theta_2 \over2 \pi }   e^{ -i \theta_1 - 2 i \theta_2} {
 1\over
(  1 -  {  e^{ - 2 i \theta_2 } \over x_2^2 } ) (1 - { e ^{ - i
\theta_1 - i \theta_2 } \over x_1 x_2 } ) } \times \\ \notag &&  ~~~~
\times  \log \Gamma(1 + i 2 g (\cos \theta_1 - \cos \theta_2 ))
\eea
This last term looks a bit messy and with many possible branch
cuts.

Let us simplify this expression a bit more. First we note that we
can write \nref{chitw} as
\bea \la{chinewxp} \chi &= & \int  { 2
e^{ - i \theta_2 }
 \over x_2 } { 1 \over 1 - { e^{ - i 2 \theta_2 } \over x_2^2 } } f(\theta_1,\theta_2)
 \\ \notag
&  = &  \int  \left[ { 1 \over
 1 - { e^{ - i \theta_2 } \over x_2 } } - { 1 \over
 1 + { e^{ - i \theta_2 } \over x_2 } } \right] f(\theta_1,\theta_2)
\\ \notag & =& \int  { 1 \over
 1 - { e^{ - i \theta_2 } \over x_2 } } \left[ f(\theta_1,\theta_2 ) - f(\theta_2 , \theta_1)
 \right]
 \\ \notag
& =&  \int  \left[ { 1 \over
 1 - { e^{ - i \theta_2 } \over x_2 } } - { 1 \over
 1 - { e^{ - i \theta_1 } \over x_2 } } \right] f(\theta_1,\theta_2)
\eea
 where we used that $f(\theta_1 + \pi , \theta_2 + \pi) =
f(\theta_2,\theta_1) $. We have also relabelled the integration
variables.

By forming the antisymmetric combination $\chi_a(1,2) \equiv
\chi(1,2) - \chi(2,1)$ it is possible to simplify this expression
further. One can see that \bea
\la{expressimp}\chi_a(1,2) &=
& i \int { d \theta_1 \over 2 \pi } { d \theta_2 \over 2 \pi } e^{
- i \theta_1 - i \theta_2 } { (x_1 - x_2)( e^{ - i \theta_2} - e^{
- i \theta_1 } ) \over ( x_1 - e^{ - i \theta_1} )( x_2 - e^{ - i
\theta_2} ) ( x_1 - e^{ - i \theta_2} )( x_2 - e^{ - i \theta_1} )
} \times \\ \notag &&~~~~~\log \Gamma(1 + i 2 g (\cos \theta_1 - \cos
\theta_2 )) \\ \notag &=& -i \oint { d z_1 \over 2 \pi } { d z_2 \over 2
\pi }   { (x_1 - x_2)( z_2 - z_1   ) \over ( x_1 -z_1 ) ( x_2 -z_2
) ( x_1 -z_2 )( x_2 -z_1 )} \times \\ \notag &&~~~~~\log \Gamma(1 + i 2 g
(\cos \theta_1 - \cos \theta_2 )) \\ \notag
 & =& -i \oint {dz_1 \over 2 \pi }  \oint { dz_2 \over 2 \pi }
\left( { 1
 \over
(x_1 - z_1) ( x_2 - z_2) } - { 1 \over (x_1 - z_2) (x_2 - z_1) }
\right)\times \\ \notag  && \times \log\Gamma( 1 + i 2 g  ( \cos \theta_1
- \cos \theta_2 ) ) \eea
 where we have set $z_i = e^{ - i
\theta_i}$. In these expressions, of course, we replace $2 \cos
\theta_1 = z_1 + { 1 \over z_1 } $, etc. Note that we can view
this last expression as arising from $\chi_a(1,2) = \tilde
\chi(1,2) - \tilde \chi(1,2)$ with $\tilde \chi$ as in \nref{integrexp},
 which is the formula we wanted to derive.

\section{Diagrammatic form of the conservation equations
in section 3}

In order to write all, a priori, possible diagrams that lead to
physical processes it is useful to adopt a diagrammatic form that
matches each set of equation one to one. If we consider a general
interaction vertex (given by the solutions to the conservation
equations (\ref{conseqd})) we see that all possible situations
(where the quantum numbers $n$ are considered free) arise from
crossing transformations of the vertices labeled $\alpha$ and
$\beta$. In total, there are six possibilities for each graph,
giving a total number of 12 vertices (of course, for given
incoming and outgoing particles of fixed $n$ there will always be
two possibilities). A convenient representation is a double line
notation for each state, where each line joining spectral
parameters represents an equality. If we follow a pair of lines
forward in time the right line always represents $X^+$ while the
left one represents $X^-$. We also allow these lines to twist
(cross), representing interchange of $+ \leftrightarrow -$ and (as
it will be convenient to consider crossing) inversion. That is, if
two spectral parameters, say $X^+$ and $Y^+$ are connected, then
$X^+=(Y^+)^{(-1)^k}$, where $k$ is the number of times that line
crossed other lines. In particular, the vertices $\alpha$ and
$\beta$ are given by

\begin{figure}[h]
         \centerline{
           \scalebox{1}{
             \input{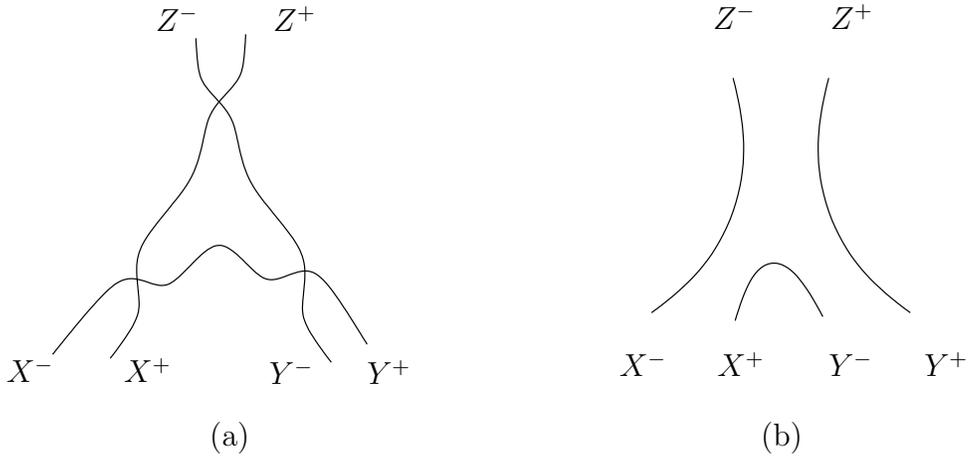}
             }
           }
         \caption{These are the vertices $\alpha$ (a) and $\beta$ (b)
          described in section 3. Each magnon is represented by a double line
         (a ``string'') that is allowed to twist. Notice that since all
         lines cross other lines twice there is no inversion to
         consider. Using crossing symmetry we can obtain all other diagrams.}
         \label{vertline}
       \end{figure}

Since one wants to draw complicated diagrams (like the ones in
section 3), the double line notation might get confusing. An
economic way of representing the same information is to consider
single line diagrams, where the single line separates $+$ from
$-$. We add a dot on top of the single line, near the vertex, when
that ``string'' is twisted. In this simple way, applying crossing
symmetry is just adding a dot (or suppressing it) and move the
particle from the past to the future (or viceversa). In this way
we can present all possible vertices as:

\begin{figure}[h]
         \centerline{
           \scalebox{1}{
             \input{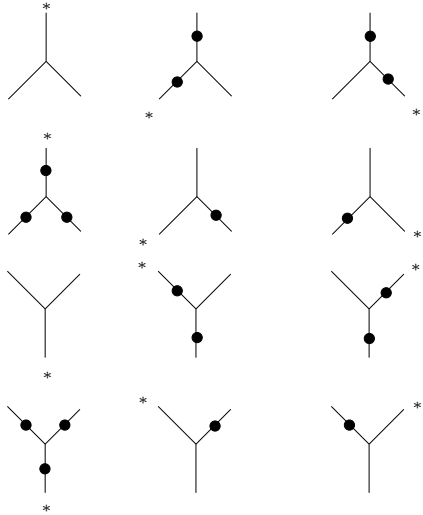}
             }
           }
         \caption{These are all possible vertices in single line notation. Vertex $\beta$ is the first in the first line, while $\alpha$ is the
         first in the second line. All other vertices come from applying crossing to these two. Notice, also, that lines 2 and 4
         are obtained from 1 and 3 by
         just adding dots where there was none and suppressing them
         where there used to be one. This amounts to changing $+
         \leftrightarrow -$. Finally the $*$ marks the leg that
         carries the higher charge $n_1+n_2$, as all vertices that we consider join
         charges, $n_1$, $n_2$ and $n_1+n_2$. As promised there is only two vertices (related by $+
         \leftrightarrow -$) for a given structure of $n$'s.}
         \label{vertdot}
       \end{figure}

From these building blocks we can construct any complicated
diagram. Notice that although there are 12 vertices, there are
only 6 types of equations in each vertex (2 for every $n$
structure). That means that we can only think of vertices when two
particles come from the past and one goes to the future (the first
six in figure \ref{vertdot}). The others just represent the same
equations (lines 2 and 3 are equivalent as are 1 and 4 in the same
figure).

If we have a fixed structure for the $n$'s carried by each line in
the graph (as in the analysis of section 3), we only have to be
careful about the $*$'s in the vertices (see caption of figure
\ref{vertdot}). One nice feature of this notation is that if we
are only interested in equations relating the spectral parameters
of external lines we can ``detach'' the dots from the vertices and
assign them to propagators. If a propagator ends up with two dots,
from both neighboring vertices, they just cancel each other. Once
this is done it is very simple to read the equations from the
diagram by following the parameters along the lines, remembering
that dots interchange $+$ and $-$ and that the final equation
contains an inverse if we went through an odd number of dots.

There are a number of rules one can derive from these graphs. For
example, if all external particles have $n=1$, then the number of
dots in the internal propagators of a box diagram is even. This is
a consequence of the fact that a box diagram always has two $2 \to
1$ vertices and two $1 \to 2$ vertices. Notice also that $n$ has
to go back to itself after going around the loop. That also means
that we have to have two vertices that increase $n$ and two that
decrease $n$.

Let us work out an example explicitly. The box diagram that was
discussed in detail in section 3 (figure \ref{Afig4}) can be drawn
in this notation as

\begin{figure}[h]
         \centerline{
           \scalebox{1}{
             \input{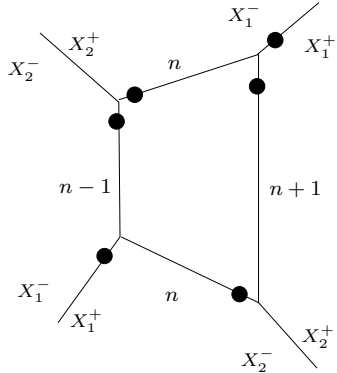}
             }
           }
         \caption{This is the ``dotted'' version of figure \ref{Afig4} in section 3.}
         \label{dotexample}
       \end{figure}

It is easy to read from this picture that the outer spectral
parameters just get mapped to themselves. The only constraint
comes from the ``mass'' equation (\ref{const}) for the propagators
in the box. From analyzing one of the propagators, for example the
bottom one, we see that this is just

\begin{eqnarray}
\left( X_2^{-}+\frac{1}{X_2^{-}}\right)-\left(
X_1^{-}+\frac{1}{X_1^{-}}\right) & = & \frac{i n}{g}
\end{eqnarray}

\noindent which leads to the condition obtained for the double
pole in section 3.

\section{Classical magnon solutions found by Spradlin and Volovich
\cite{sv}}\label{svsolutions}

For the reader's convenience we reproduce the solutions in
\cite{sv}. We will use complex target space coordinates $Z_1=X_1+
i X_2$ and $Z_2 = X_3 + i X_4$  describing an $S^3$ given by
$|Z_1|^2+|Z_2|^2=1$. The one magnon solution, written in terms of
the $\lambda^\pm$ and worldsheet $\sigma^{0,1}$ variables, is
given by
\begin{eqnarray}
Z_1 &=& \frac{e^{+i \sigma^0}}{\sqrt{\lambda^+ \lambda^-}}
\frac{\lambda^+ e^{-2 i \mathcal{Z}^+}+\lambda^- e^{-2 i
\mathcal{Z}^-}}{e^{-2 i
\mathcal{Z}^+}+ e^{-2 i \mathcal{Z}^-}}\\
Z_2 &=& \frac{e^{-i \sigma^0}}{\sqrt{\lambda^+ \lambda^-}} \frac{i
(\lambda^- - \lambda^+)}{e^{-2 i \mathcal{Z}^+}+ e^{-2 i
\mathcal{Z}^-}}
\end{eqnarray}
\noindent where we defined
\begin{equation}
\mathcal{Z}^\pm =
\frac{1}{2}\left(\frac{\sigma^1-\sigma^0}{\lambda^\pm-1} +
\frac{\sigma^1+\sigma^0}{\lambda^\pm +1}\right)
\end{equation}
A two magnon (scattering) solution was also presented in \cite{sv}
and discussed in section 4. The form of this solution is
\begin{eqnarray}
Z_1=\frac{e^{i
\sigma^0}}{2\sqrt{\lambda_1^+\lambda_2^+\lambda_1^-\lambda_2^-}}
\frac{N_1}{D}\\
Z_2=\frac{-
i}{2\sqrt{\lambda_1^+\lambda_2^+\lambda_1^-\lambda_2^-}}\frac{N_2}{D}
\end{eqnarray}
\noindent with
\begin{eqnarray}
D &=&\lambda_{12}^{++}\lambda_{12}^{--}\cosh(u_1+u_2) +
\lambda_{12}^{+-}\lambda_{12}^{-+} \cosh(u_1-u_2) + \nonumber\\
& & \lambda_{11}^{+-}\lambda_{22}^{+-}\cos (v_1-v_2)\\
N_1 &=&
\lambda_{12}^{++}\lambda_{12}^{--}\left[\lambda_1^+\lambda_2^+
e^{+u_1+u_2}+\lambda_1^-\lambda_2^- e^{-u_1-u_2}\right] +
\nonumber\\
& &\lambda_{12}^{-+}\lambda_{12}^{+-}\left[\lambda_1^+\lambda_2^-
e^{+u_1-u_2}+\lambda_1^-\lambda_2^+ e^{-u_1+u_2}\right] + \nonumber\\
& & \lambda_1^+\lambda_1^-\lambda_{11}^{+-}\lambda_{22}^{+-} e^{i
(v_1-v_2)} +
\lambda_2^+\lambda_2^-\lambda_{11}^{+-}\lambda_{22}^{+-} e^{-i
(v_1-v_2)}\\
N_2 &=& \lambda_{11}^{+-} e^{i v_1}
\left[\lambda_{12}^{++}\lambda_{12}^{-+}\lambda_2^- e^{u_2} +
\lambda_{12}^{--}\lambda_{12}^{+-}\lambda_2^+ e^{-u_2}\right] +
\nonumber\\
& & \lambda_{22}^{+-} e^{i v_2}
\left[\lambda_{21}^{++}\lambda_{21}^{-+}\lambda_1^- e^{u_1} +
\lambda_{21}^{--}\lambda_{21}^{+-}\lambda_1^+ e^{-u_1}\right]
\end{eqnarray}

\noindent and

\begin{eqnarray}
\lambda_{jk}^{\pm\pm} &=& \lambda_j^\pm - \lambda_k^\pm\\
u_j &=& i \left[\mathcal{Z}^+_j-\mathcal{Z}^-_j\right]\\
v_j &=& \mathcal{Z}^+_j+\mathcal{Z}^-_j - \sigma^0
\end{eqnarray}

These solutions admit the following generalization: $u_j
\rightarrow u_j + a_j$ and $v_j \rightarrow v_j + b_j$, with $a_j$
and $b_j$ real. Two of these four parameter can be reabsorbed by a
worldsheet coordinate redefinition. We are, therefore, left with
two parameters corresponding to a relative distance and a relative
phase.

The scattering solution presents the symmetry $\lambda_1^+
\leftrightarrow \lambda_2^+$ or, alternatively  $\lambda_1^-
\leftrightarrow \lambda_2^-$. Also, it collapses to the single
magnon solution for $\lambda_j^+=\lambda_j^-$ for $j= $ 1 or 2.

\end{document}